\documentclass{article}

\usepackage{arxiv}

\usepackage[utf8]{inputenc} 
\usepackage[T1]{fontenc}    

\usepackage{url}            
\usepackage{booktabs}       
\usepackage{amsfonts}       
\usepackage{nicefrac}       
\usepackage{lipsum}		
\usepackage{graphicx}
\usepackage{natbib}
\usepackage{doi}
\usepackage[utf8]{inputenc}
\usepackage[english]{babel}
\usepackage[T1]{fontenc}
\usepackage{amsmath}
\usepackage{amsfonts}
\usepackage{graphicx}
\usepackage{amssymb}
\usepackage{algorithm}
\usepackage{subcaption}
\usepackage{stackengine}
\usepackage[final]{pdfpages}
\usepackage{makecell}
\usepackage{longtable}
\usepackage{tikz}
\usepackage[utf8]{inputenc}
\usepackage[english]{babel}
\usepackage[T1]{fontenc}
\usepackage{amsmath}
\usepackage{amsfonts}
\usepackage{graphicx}
\usepackage{amssymb}
\usepackage{algorithm}
\usepackage{subcaption}
\usepackage{stackengine}
\usepackage[final]{pdfpages}
\usepackage{makecell}
\usepackage{longtable}
\usepackage{tikz}
\definecolor{ao(english)}{rgb}{0.0, 0.5, 0.0}
\definecolor{forestgreen}{RGB}{34,139,34}
\definecolor{deepskyblue}{RGB}{0,0,205}
\usetikzlibrary{calc, automata,positioning, decorations.pathreplacing}
\usepackage{booktabs}
\usepackage{multirow}

\usepackage{standalone}
\usepackage{siunitx}
\usepackage{natbib}
\setcitestyle{numbers}
\setcitestyle{square}

\usetikzlibrary{backgrounds,calc,shadings,shapes.arrows,shapes.symbols,shadows}
\usetikzlibrary{positioning}

\usepackage{mathtools}
\usepackage{algorithmicx}
\usepackage{algpseudocode}
\usepackage{color}
\definecolor{red}{rgb}{1,0,0}
\definecolor{blue}{rgb}{0,0,1}
\definecolor{green}{rgb}{0,1,0}
\newcommand{\kibitz}[2]{\ifnum\Comments=1\textcolor{#1}{#2}\fi}

\DeclareMathOperator*{\argmax}{arg\,max}

\usepackage{alphalph}
\usepackage{optidef}
\usepackage{etoolbox}
\usepackage{eurosym}

\AtBeginDocument{%
  \AtBeginEnvironment{mini!}{%
  }
}

\algnewcommand\algorithmicinput{\textbf{Input:}}
\algnewcommand\Input{\item[\algorithmicinput]}
\algnewcommand\algorithmicoutput{\textbf{Output:}}
\algnewcommand\Output{\item[\algorithmicoutput]}
\usepackage[english]{babel}

\algnewcommand\algorithmicforeach{\textbf{for each}}
\algdef{S}[FOR]{ForEach}[1]{\algorithmicforeach\ #1\ \algorithmicdo}
\definecolor{yellowrecpol}{HTML}{CCCC00}

\title{Optimal Control of Renewable Energy Communities subject to Network Peak Fees with Model Predictive Control and Reinforcement Learning Algorithms}

\date{} 					

\author{ Samy Aittahar \\
	Department of Computer Science\\
	University of Liège\\
	Liège, Belgium \\
	\texttt{saittahar@uliege.be} \\
	\And
	Adrien Bolland \\
	Department of Computer Science\\
	University of Liège\\
	Liège, Belgium \\
    \And
	Guillaume Derval \\
	Department of Computer Science\\
	University of Liège\\
	Liège, Belgium \\
    \And
	Damien Ernst \\
	Department of Computer Science\\
	University of Liège\\
	Liège, Belgium \\
}

\renewcommand{\headeright}{Preprint. Under review.}
\renewcommand{\undertitle}{}
\renewcommand{\shorttitle}{}

\hypersetup{
pdftitle={A template for the arxiv style},
pdfsubject={energy},
pdfauthor={Samy Aittahar, Adrien Bolland, Guillaume Derval, Damien Ernst},
}

\begin{document}
\maketitle

\begin{abstract}
	We propose in this paper an optimal control framework for renewable energy communities (RECs) equipped with controllable assets. Such RECs allow its members to exchange electricity production surplus through an internal market. The objective is to control their assets in order to minimise the sum payable of individual electricity bills. These bills account for the electricity exchanged through the REC and with the retailers. Typically, for large companies, another important part of the bills is the costs related to power peaks; in our framework, they are determined from the energy exchanges with the retailers. We compare rule-based control strategies with the two following control algorithms. The first one is derived from model predictive control techniques, and the second one is built with reinforcement learning techniques. We also compare variants of these algorithms that ignore peak power costs. Results confirm that using policies which account for the power peaks leads to significantly lower sum payable for individual electricity bills and thus better control strategies at the cost of higher computation time. Furthermore, policies trained with reinforcement learning approaches appear promising for real-time control of the communities, where model predictive control policies may be computationally expensive in practice. These findings encourage the pursuit of the efforts toward development of scalable control algorithms, operating from a centralised standpoint, for renewable energy communities equipped with controllable assets.
\end{abstract}


\section{Introduction}
\label{intro}
Decentralisation of renewable electricity production, close to its place of consumption, is gaining increasing attention as a promising approach to decarbonise the demand of electricity \cite{SOEIRO2020e04511,DISILVESTRE2021111565, sudhoff2022operating}. In this context, the European Union (EU) has provided a legal context introducing \emph{renewable energy communities} \cite{europeanrules}. A renewable energy community (REC) is a local electricity market where the members, equipped with renewable energy generation and storage assets, can share local and decarbonised electricity production with the other members; we refer to this aggregated electricity production as the \emph{REC production}. In practice, the members have joined the REC in order to make cost savings in their electricity bills, where they are charged by their retailers proportionally to their electricity consumption. Moreover, in many parts of the EU, for large companies, the electricity bills also includes another (but prominent) grid fee related to their consumption and production peaks in the main electrical grid, which we refer to as \emph{offtake peaks} and \emph{injection peaks}, respectively. These large companies thus expect to significantly maximise their cost savings by joining an REC. This can be achieved by efficiently managing the controllable assets (e.g., batteries) within the REC; we explore this decision-making issue in this paper.

According to European directives, the local market clearing process of RECs is performed at fixed periods, which we refer to as \emph{market periods}, through the reallocation of the REC production by a central entity, namely the Energy Community Manager (ECM). Among their responsibilities, the ECM must communicate for longer periods (e.g., every month) the results of the clearings performed during this period to the Distribution System Operator (DSO). We refer to the latter as the \emph{billing period}. The ECM must ensure the compliance of the reallocation of the REC production with the regional (or national) regulations,, which notably requires that the REC production is fully allocated to the members within the limits of the amount of energy they have consumed, and that no energy is bought from the retailers to be sold to a member through the REC (and vice-versa). Once the DSO has accepted the clearing results (in keeping with the regulation), they transfer them to the retailers to compute the electricity bills of each REC member, accounting for the reallocation of the REC production. We refer to these electricity bills as \emph{ex-post} electricity bills. The ECM is often remunerated by a fraction of the cost savings realised in the \emph{ex-post} electricity bills of the members. We thus assume, in this paper, that the ECM aims to minimise the sum payable of the \emph{ex-post} electricity bills; note that we ignore the remuneration of the ECM in these bills. We refer to this minimised sum payable of \emph{ex-post} electricity bills as the \emph{global REC bill}.


\emph{Ex-post} electricity bills include costs related to electricity exchanges with the retailers and through the REC (as determined by the ECM for each market period). On the one hand, the tariffs of buying electricity from the retailers mainly include the price of the energy consumed, the distribution and transmission fees (the tariffs related to these fees are fixed by the DSO for all members), and other taxes; for prosumers, the electricity sold to the retailers, at a price fixed by contract, is deducted from the electricity bills. On the other hand, the electricity production shared through the REC is only subject to distribution fees; we note that they differ from the fees fixed by the DSO. As mentioned above, grid fees related to offtake and injection peaks are also part of these \emph{ex-post} electricity bills. However, the European regulation does not specify whether these grid fees should be computed as if no REC has been implemented. We argue that the injection and offtake peaks should only be computed from the energy exchanged with the retailers in order to accelerate the development of RECs; this is also an assumption that we will make throughout this paper. Indeed, the overall REC electricity consumption and production are physically seen by the main electrical grid as already aggregated. Furthermore, these costs often constitute a significant portion of the electricity bills. Beyond the decarbonation of the demand and the cost savings in the electricity bills, investing into assets that foster consumption from local renewable sources reduces the risks of outages of the main electrical grid, particularly in the winter \cite{LALEMAN2016416, en12040682}. We note however that the optimal control framework proposed in this paper can be easily adapted to the case where power peaks costs would be computed without accounting for the REC.


The scientific literature, discussed in Section \ref{relwork}, does not provide, to date and to the best of our knowledge, any existing modelling framework that explicitly optimise the controllable assets usage in renewable energy communities towards the minimisation of global REC bills over time. To address this gap, we propose in this paper a framework for modelling the optimal usage of controllable assets in RECs. In Section \ref{probstat}, we describe the problem of acting on the controllable assets to minimise the sum payable of global REC bills over time as a Partially Observable Markov Decision Process (POMDP). The dynamics of this POMDP represent the electricity power flows of the members, which are influenced by external events, and the reward function reflects the (negated) global REC bill as computed by the ECM. We describe, in Section \ref{policies}, two practical control algorithms that aim to minimise the sum payable of the global REC bills over time. The first one follows a \emph{model predictive control} scheme \cite{rawlings2017model} by solving an internal optimisation problem built from the POMDP specifications and predictions of incoming external events. The second one is built through a \emph{reinforcement learning} scheme, resulting from a simulation-based policy-search procedure that seeks to approximate the optimal policy \cite{sutton2018reinforcement, sutton1999policy}. We also introduce variants of these policies that ignore the costs related to the peaks. In Section \ref{testpolicies}, we compare these policies with rule-based strategies against RECs with two members built from synthetic data and with seven members derived from historical data of an existing REC located in Belgium. We conclude this paper in Section \ref{conclusion} with a detailed discussion on recommendations of research directions to pursue towards the development of scalable control algorithms dedicated to RECs.

\section{Related work}
\label{relwork}

The work carried out in this paper is related to decision-making issues in REC. We divide this related work into three parts. The first part is related to the main generic decision-making problems associated with RECs. The second part focuses on model predictive control (MPC) techniques \cite{rawlings2017model} and reinforcement learning (RL) techniques \cite{sutton2018reinforcement} in microgrids, which can be seen as single-entity RECs. The third part focuses on the scarce literature of the applications of these techniques in renewable energy communities, highlighting their limitations with respect to the decision-making problem we address in this paper.

\paragraph{Decision-making issues for RECs.} In a broader perspective, many decision-making problems related to RECs are challenging. There is a particular focus on investment strategies. These strategies are influenced by energy costs, which must be carefully set to foster investments into renewable energy generation as well as devices helping to increase the self-consumption rate (e.g., storage devices)\cite{LI2023120706, LI2022104710, WUSTENHAGEN20121, belmarmodellingrec}. In \cite{en15155468}, the authors show that maximising the benefits of an REC operation does not necessarily lead to peak reductions in the main electrical grid. However, they show that fostering investments towards storage devices helps to sensitively decrease the load consumed from the main electrical grid. In \cite{manueldevillena2020allocation}, the authors introduce the concept of \emph{repartition keys} to reallocate the REC production to the members. Altogether, the results of all these works advocate for the pursuit of the efforts towards solving these decision-making challenges to support the deployment of the RECs around the world.

\paragraph{MPC and RL for microgrids.} Control algorithms derived from MPC and RL classes of techniques have been widely used in the past to operate microgrids (see review in \cite{ernstmpc}). These control algorithms are still popular due to their promising performances as well as their simplicity to implement and deploy in practice \cite{en16134851, yang2020reinforcement, rl_microgrids_survey}. In \cite{mpcvsrlpvbattery}, two algorithms derived from MPC and RL were tested on a single building with PV panels and a battery, with similar performances reported in the results, even though the authors noted that MPC algorithms are more suitable since they can be adapted more quickly for a new building configuration. In \cite{hierarchicalmpc}, the authors propose a hierarchical model predictive control scheme to control a microgrid equipped with wind turbines, PV panels and batteries in a way that simultaneously minimise the electricity bills and maximise the lifetime of the equipped devices. In \cite{deeprlmicrogridflexible} and \cite{anotherdeeprlmicrogridflexible}, the authors compare several RL algorithms against instances of microgrids and show they are able to extract near-optimal policies. Many of these works share similarities with the optimal REC control framework proposed in this paper.

\paragraph{MPC and RL for REC.} At the best of our knowledge, the literature about the design of control algorithms specifically dedicated to renewable energy communities is rather scarce. In \cite{rec_paper}, various MPC strategies have been tested on RECs. However, unlike the framework proposed in this paper, the network peak fees are not taken into account. In \cite{marlpeakrebounds}, a multi-agent reinforcement learning (MARL) algorithm is designed to optimise the controllable assets of an REC, and is compared to an algorithm combining MPC and supervised learning. Their approach allowed one to decrease energy bills as well as mitigate offtake and injection peaks. However, their results are restricted by the specific energy community structure (the members cannot possess their own individual storage devices). In \cite{zeroenergymultiagent}, the authors propose another MARL approach to optimise the balance between consumption and production, thus ignoring the overall electricity bill. In \cite{TOMIN2022903}, the authors propose a framework, based on bi-level optimisation, where peak demand and energy bills are jointly optimised with controllable assets for a single billing period. In this framework, the peaks are computed \emph{before} the reallocation of the REC production.

\section{Optimal REC Control Problem}
\label{probstat}

This section details the optimal control framework of RECs with some members owning controllable assets. In these RECs, the electricity bill of each member is the sum payable of the fees related to the exchanges through the REC, the energy bought and sold with prices fixed by the retailers, and grid costs related to their power peaks. The sum payable for these electricity bills, that is minimised by the ECM through the determination of the energy exchanged through the REC, forms the \emph{global REC bill},. In this section, we describe how an REC is structured and managed through time by the ECM. Notably, the latter periodically clears the local market through the reallocation of the REC production by following a repartition scheme that is explained in Section \ref{opt_realloc_scheme}. In practice, the ECM is remunerated by charging members either through fixed fees (e.g., annual subscription plans) or proportionally to the amount of energy exchanged through the REC. We leave the business model of the ECM out of the scope of this paper. However, the repartition scheme that we describe further in this section can easily be adapted to account for the ECM remuneration. We conclude that section by providing a full formulation of the dynamical system as a (particular case of) Partially Observable Markov Decision Process (POMDP), notably providing a time discretisation that accounts for the billing or market periods, as well as smaller time periods for a more granular usage of the controllable assets.

\subsection{REC Structure and Management}

A member of the REC is characterised by (i) its tariffs for energy bought (sold) from (to) the retailers and (ii) its means of consumption and electricity production (some of them being controllable). The tariffs related to energy exchanges with the retailers are composed of the price of (bought or sold) energy, distribution and tranmission fees, and taxes; recall that we do not consider the fixed terms in these tariffs (e.g., subscription plans). According to the contract fixed by retailers with the members (and more generally with end users), the price of energy itself is either fixed or variable through time (e.g., peak and off-peak hours or real-time markets such as BELPEX for Belgium). To maintain a reasonable complexity in this paper, we only consider contracts with fixed-price plans. However, the framework that we propose in this paper may easily be adapted for variable prices of energy through time. Energy exchanges through the REC are subject to distribution fees that are proportional to the amount of energy exchanged. Unlike the tariffs of energy exchanges with retailers, the tariffs related to distribution fees within the REC are fixed for all members. Note that these distribution fees, to which we concisely refer in the remainder of this paper as \emph{REC fees}, are different than those that are charged by the DSO (to the users of its electrical network) for electricity exchanges with the retailers. Additional (but often prominent) costs are grid fees related to offtake and injection peaks. Similarly to the REC fees, their respective tariffs are fixed for all the members by the DSO. In this paper, we assume that these peaks correspond, for each billing period and for each member of the REC, to maximum values of energy exchanged with the retailers across the market periods elapsed in that billing period. We also assume that peak tariffs are the highest in the electricity bills of the members of the REC; however, this is not a strict hypothesis, but rather a design choice that follows the typical structure of electricity bills.

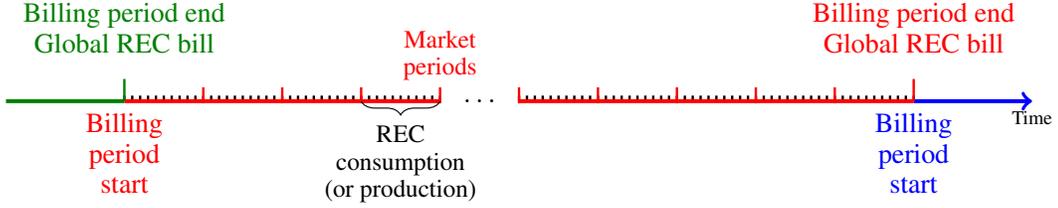
\begin{figure}
  \begin{tikzpicture}
    every node/.style={
        font=\scriptsize,
        text height=1ex,
        text depth=.25ex,
    },
]
\draw[ao(english), -, line width=0.5mm] (-1.5,0) -- (0,0);

\foreach \x in {0,...,40}{
    \draw[black, line width=0.3mm] (\x*0.1 cm,2pt) -- (\x*0.1 cm,0pt);
}
\foreach \x in {50,...,100}{
    \draw[black, line width=0.3mm] (\x*0.1 cm,2pt) -- (\x*0.1 cm,0pt);
}

\foreach \x in {1,...,9}{
    \draw[red, line width=0.3mm] (\x cm,5pt) -- (\x cm,-0.8pt);
}

\draw[red, -, line width=0.5mm] (0,0) -- (4,0);
\draw[red, -, line width=0.5mm] (5,0) -- (10,0);
\draw[blue, ->, line width=0.5mm] (10.0000001,0) -- (11.5,0);

\draw[ao(english), line width=0.3mm] (0 cm,8pt) -- (0 cm,0pt);
\draw[red, line width=0.3mm] (10 cm,8pt) -- (10 cm,0pt);

\node[anchor=north] at (0,0) {\color{red}\makecell{Billing\\ period\\ start}};
\node[anchor=north] at (1,0) {};
\node[anchor=north] at (2,0) {};
\node[anchor=north] at (3,0) {};
\node[anchor=north] at (4,0) {};
\node[anchor=north] at (5,0) {};
\node[anchor=north] at (4.5,0.15) {$\ldots$};
\node[anchor=north] at (6,0) {};
\node[anchor=north] at (7,0) {};
\node[anchor=north] at (8,0) {};
\node[anchor=north] at (9,0) {};
\node[anchor=north] at (10,0) {\color{blue}\makecell{Billing\\ period\\ start}};
\node[anchor=north] at (11.5,0) {\scriptsize Time};

\draw 
    node[anchor=north,midway,above=14pt] {\parbox{5.5cm} {\centering \color{ao(english)}{{Billing period end}}\\ {Global REC bill}}};
\draw 
    node[anchor=north,midway,above=4pt, xshift=4cm] {\parbox{5.5cm} {\centering \footnotesize \color{red}{{Market}}\\ {periods}}};
\draw 
    node[anchor=north,midway,above=14pt, xshift=10cm] { \parbox{5.5cm} {\centering {\color{red}{Billing period end}}\\ {\color{red}{Global REC bill}}}};
\draw[decorate,decoration={brace,amplitude=5pt}] (4,0) -- (3,0)
    node[anchor=north,midway,below=5pt] { \parbox{4.5cm} {\centering \footnotesize REC \\ consumption \\ (or production)}};
\end{tikzpicture}
  \caption{Renewable energy community timeline. During a billing period, each member consumes and produces in real time electricity on their own and by the usage of their controllable assets; in the timeline, black ticks refer to time steps during which control actions are taken. At the end of a billing period, the ECM computes the (optimal) reallocation of the REC production for each market period. This reallocation creates new meter readings and these are passed on by the ECM to the DSO and the retailers to compute the global REC bill.}
  \label{fig:timeline}
\end{figure}

The net consumption and production of the members (i.e., after accounting for self-consumption at electrical level) are independently metered in real time. At the end of fixed time periods, that we denote as \emph{billing periods} (e.g., every month), we assume that the ECM clears the local market by following a known reallocation scheme, which is described in Section \ref{opt_realloc_scheme}, to share the REC production to the other members for each market period. This reallocation scheme aims to minimise the sum payable of the \emph{ex-post} electricity bills while ensuring its compliance with the European Union directives. The result of this reallocation scheme, which may be concisely represented as \emph{repartition keys} \cite{manueldevillena2020allocation}, is passed on by the ECM to the retailers and the DSO (along with the meter readings). Once the reallocation scheme is accepted by the DSO, the retailers pass on the ex-post electricity bills to their respective members. These bills account for the energy bought (sold) from (to) the retailers, the distribution fees related to the energy exchanges through the REC, the costs related to the offtake and injection peaks (accounting for the reallocation scheme). Figure \ref{fig:timeline} illustrates the timeline of the REC. 




\subsection{Optimal reallocation scheme}
\label{opt_realloc_scheme}
For a member, sharing a fraction of its energy production or receiving a fraction of the REC production impacts its ex-post electricity bill. More precisely, if a member has a surplus of electricity production while another has a positive net consumption, selling the former to the retailer and buying the latter from another retailer will cost more than sharing the production via the REC, notably because of the peaks measured after sharing the REC production (retailer's/main grid point of view). We assume that the ECM applies an \emph{optimal reallocation} scheme. This scheme, resulting in the global REC bill, is formalised hereafter.

Let us consider an REC with $M$ members for which the electricity bills are computed for $R$ market periods within a billing period. Let $C^-_{m, r}$ and $C^+_{m, r}$ be the values of the meter readings, both non-negative and expressed in \si{kWh}, of a member $m \in \left\{ 1, \ldots, M\right\}$ that respectively correspond to its net consumption and its net production for the market period $r \in \left\{ 1, \ldots, R\right\}$. Traditionally, this member is billed by its retailer, for the entire billing period, at a cost of $B^-_m$ for its net consumption and remunerated at a price of $B^+_m$ for its net production. Recall that these cost coefficients are sums of proportional terms (to energy amounts, we ignore fixed terms) related to prices of buying and selling energy, to distribution/transmission fees, and to other taxes. Electricity flows create offtake and injection peaks, which respectively correspond to the greatest values of the net consumption and the net production meter readings measured during the billing period. These values are billed to the members through the unique tariffs $P^-$ and $P^+$ that are applied to offtake and injection peaks, respectively. In this paper, we do not convert the peak values into power units (\si{kW}) to avoid overburdening the reallocation scheme modelling. If the member was not participating in the REC, its electricity bill, to which we refer as $EB_m$, would be computed as follows:

\begingroup
    \small%
\begin{equation}
\label{eq:elec_bill_comput}
EB_m = \left( \sum_{r = 1}^{R} B_{m}^- C_{m, r}^- - B_{m}^+ C_{m, r}^+ \right) + P^-\max_{r \in \left\{1, \ldots, R\right\}}{C^-_{m, r}} + P^+\max_{r \in \left\{1, \ldots, R\right\}}{C^+_{m, r}}.
\end{equation}
\endgroup

In other words, energy consumption and production are proportionally billed and remunerated at the prices fixed by the retailers for each market period and peaks are proportionally billed at unique prices for all members, We denote as $EB = \sum^M_1 EB_m$ the total combined value of the electricity bills of the members. Inside an REC, the members can exchange their electricity production (with the other members) at each market period; these exchanges, characterised by either sharing a fraction of the net production of a member with the REC or sharing a fraction of the REC production with a member, are subject to REC fees that are defined by $\Lambda^+$ and $\Lambda^-$, respectively. Note that these distribution fees are not necessarily equal to the hidden distribution fees that fall within prices fixed by the retailers. To model the energy exchanges through the REC, we introduce the decision variable $e^-_{m, r}$, which is the quantity of electricity production allocated to the member $m$ for the market period $r$. Similarly, we introduce the decision variable $e^+_{m, r}$ which is the quantity of electricity production shared by the member $m$ for the market period $r$. Recall that the ECM cannot shape its repartition scheme in order to buy energy from the retailer of a member to share it with another member through the REC. To that end, the value of the former ($e^-_{m, r}$) is bounded by the value of the consumption meter reading after accounting of the production meter reading, and vice versa for the latter ($e^+_{m, r}$):

\begin{align}
    \label{eq:constrrealloca_1}
    0 \leqslant e_{m, r}^- &\leqslant \max(C_{m, r}^- - C_{m, r}^+, 0), \; \forall m \in  \left\{1,\ldots,M\right\}, \; \forall r \in  \left\{1,\ldots,R\right\},\\
 \label{eq:constrrealloca_2}
    0 \leqslant e_{m, r}^+ &\leqslant \max(C_{m, r}^+ - C_{m, r}^-, 0), \; \forall m \in  \left\{1,\ldots,M\right\}, \; \forall r \in  \left\{1,\ldots,R\right\}.
\end{align}

\noindent Any quantity of net electricity production exchanged through the REC is shared with a member:

\begin{equation}
\label{eq:constrrealloca_3}
\sum_{m=1}^{M} e_{m, r}^- = \sum_{m=1}^{M} e_{m, r}^+, \; \forall r \in  \left\{1,\ldots,R\right\}.
\end{equation}

Within the REC, the offtake and injection peaks are measured after the reallocation of the REC production, and the energy exchanges in the REC are still subject to the network fees. Once the REC production has been reallocated (by the ECM), the ex-post electricity bill of the current billing period is computed (at the end of this billing period) for a given member as follows:

\begin{align}
\label{eq:objreallocorig}
\nonumber &EB'_m(e^-_m, e^+_m) = \\[-1em] \nonumber \\ &\left( \sum_{r=1}^{R}{B_{m}^- (C_{m, r}^- - e_{m, r}^-) - B_{m}^+ (C_{m, r}^+ - e_{m, r}^+) + \Lambda^- e_{m, r}^- + \Lambda^+ e_{m, r}^+} \right) \\[0.5em]
\nonumber &+ P^- \left(\max_{r \in \left\{1, \ldots, R\right\}}{C_{m, r}^- - e_{m, r}^-}\right) + P^+ \left( \max_{r \in \left\{1, \ldots, R\right\}}{C_{m, r}^+ - e_{m, r}^+} \right),
\end{align}

\noindent where $e^-_m = (e^-_{m, 1}, \ldots, e^-_{m, R})$, $e^+_m = (e^+_{m, 1}, \ldots, e^+_{m, R})$. Note that intermediate ex-post electricity bills can be computed at any time during the billing period (accounting for the elapsed time within that billing period). More formally, let $\tau \in ]0; 1]$ be a ratio value of the time elapsed in the billing period; $\tau = 1$ corresponds to a fully elapsed billing period (in which case ex-post electricity bills are computed by Equation \eqref{eq:objreallocorig}). If $\tau < 1$, we compute, for each member, its intermediate ex-post electricity bill as follows:

\begin{align}
\label{eq:objrealloc}
\nonumber &EB''_m(e^-_m, e^+_m, \tau) = \\[0.5em]  &\left( \sum_{r=1}^{\lceil\tau R \rceil}{B_{m}^- (C_{m, r}^- - e_{m, r}^-) - B_{m}^+ (C_{m, r}^+ - e_{m, r}^+) + \Lambda^- e_{m, r}^- \Lambda^+ e_{m, r}^+} \right) + \\[0.5em]
\nonumber &+ \tau \left[ P^- \left(\max_{r \in \left\{1, \ldots, \lceil\tau R \rceil\right\}}{C_{m, r}^- - e_{m, r}^-}\right) + P^+ \left( \max_{r \in \left\{1, \ldots, \lceil\tau R \rceil\right\}}{C_{m, r}^+ - e_{m, r}^+} \right) \right].
\end{align} \noindent Note that, if $\tau$ < 1, values of meter reading inputs for all members ($C^+$ and $C^-$) that comes after the market period indexed by $\lceil\tau R \rceil$ are undefined, and so are corresponding variables ($e^+$ and $e^-$).

The goal of the ECM is to minimise the sum payable of ex-post electricity bills by identifying the optimal energy exchanges through the REC. The (possibly intermediate) global REC bill, accounting for the elapsed time during the billing period and referred to as $GRB_\tau$, is the result of the following minimisation problem:

\begin{equation}
\label{eq:objreallocaoptim}
GRB_\tau = \min_{(e^{*, -}, e^{*, +})} \sum^M_{m=1} EB'_m(e^{*, -}_m, e^{*, +}_m, \tau),
\end{equation}
\noindent where $e^{*, -} = (e^{*, -}_1, \ldots, e^{*, -}_M)$ and $e^{*, +} = (e^{*, +}_1, \ldots, e^{*, +}_M)$. We provide, in Appendix \ref{opt_scheme_example}, illustrative examples of this optimal reallocation scheme in order to provide some intuition about this formulation and its complexities and to show the importance of accounting for the peak costs during the computation of the optimal reallocation scheme.

\subsection{Decision process associated with RECs}
\label{pomdp_descr}
We formalise the above-mentioned REC dynamical system as an infinite-time decision process. The latter is a particular instance of Partially Observable Markov Decision Processes (POMDP) in that the agent interacting with this decision process fully observes the electrical state of the system (in this case, controllable assets state and meter readings) but has a partial view on external events that are not influenced by its decision. However, these external events have an impact, altogether with the decisions of the agent, on the state transitions (meter readings increments) and reward function (the global REC bill). We recall that a billing period covers a fixed number of market periods. Typically, market periods have a fixed duration (e.g., fifteen minutes). During a market period, there is a need to take actions on the controllable assets. Thus, we introduce a time discretisation inside a market period such that, at any time step, an action can be applied to controllable assets for a fixed duration. More formally, this POMDP provides a time discretisation from a time step ($t \in \mathcal{T} = \mathbb{N}$) to the next, which also accounts for the duration of market periods and billing periods, respectively expressed as a number of time steps and a number of market periods. We provide below the components of this decision process that we describe from a high-level perspective (see Appendix \ref{mdpmath} for more mathematical details):


\paragraph{State space} A state of the system contains, for each member, the current state of their controllable assets, the elapsed time of the current billing period and their meter readings, collected at the end of each elapsed market period (including the current one) for the current billing period. We denote as $\mathcal{S}$ the set of all such states, and the unique initial state as $S_0 \in \mathcal{S}$.

\paragraph{Exogenous space} An exogenous variable contains, for each member, their consumption and production powers generated from their non-controllable assets, averaged during the current time interval. We denote as $\mathcal{E}$ the set of such exogenous variables. We assume that the initial exogenous variable is a random variable following a probability distribution $P^{\mathcal{E}}_0(\cdot)$ and that exogenous variables transitions from any time step $t$ to the next are steered by an unknown non-Markovian conditional distribution that we denote as $P^{\mathcal{E}}(e_{t+1}|e_{0:t})$ for all $t \in \mathcal{T}$, where $e_{0:t} = (e_0, \ldots, e_t)$.

\paragraph{Action space} An action contains the (control) decisions to be applied to controllable assets. We denote the set of such actions as $\mathcal{U}$. The set of actions that can be applied in a given state $s \in \mathcal{S}$ may be restricted, and are referred to as \emph{admissible actions}. These admissible actions are given by the function $U(s)$, such that $U(s) \subseteq \mathcal{U}$.

\paragraph{Transition dynamics} The transition dynamics of the state space are defined as follows. Controllable assets states are updated for each discrete time step $t \in \mathcal{T}$ according to a model depending on the asset specifications, their current state $s_t \in \mathcal{S}$ and the action $u_t \in U(s_t)$. After the electricity flows have been computed by accounting for the controllable assets usage, the meter readings are updated for the market period to which the next time step is associated. The state transition is concisely expressed by the function $f$ as follows: 
\begin{equation}
s_{t+1} = f(s_t, e_t, u_t).
\end{equation} 

\paragraph{Cost function} The cost function describes the global REC bill computation as described in Section \ref{opt_realloc_scheme}. More formally, this function provides a zero cost at all time steps except at the end of each billing period where the solution $GRB$ of the optimal reallocation scheme is returned. In short, the cost signal is defined as follows:

\begin{equation}
    \rho_t = \begin{cases}
        GRB_1 & \text{if } t \text{ is the last timestep of a billing period,}\\
        0 & \text{otherwise.}
    \end{cases}
\end{equation} 
\noindent Note that this cost signal can be easily modified to output intermediate global REC bills at other time steps:
\begin{equation}
\label{}
    \rho_t = \begin{cases}
        GRB_{\tau(t)} & \text{\makecell{if an intermediate global REC bill\\ \hspace{-1.95cm} is needed at time step $t$,}}\\
        0 & \text{otherwise,}
    \end{cases}
\end{equation}
\noindent where $\tau(t)$ is the value of $\tau$ in Equation \eqref{eq:objrealloc} that is computed from (the information contained in) state $s_t$.

To maintain a reasonable complexity, we ignore the operational costs related to the controllable assets (e.g., discharging fees for storage devices). In practice, the cost function can be extended to incorporate these operational costs at each time step depending on the actions performed.

\section{Optimal control of RECs}
\label{policies}
A process that chooses the next action to execute based on the current information of the system is called a \emph{policy}. This information is composed of the current state and the history of past exogenous variable values. In this section, we describe how to select a policy that optimally reduces the sum payable of the global REC bills of the REC through time, as follows. Since these optimal policies cannot be computed directly as $P_0^{\mathcal{E}}$ and $P^{\mathcal{E}}$ are not known, we propose to approximate them through practical policies derived from model predictive control and reinforcement learning schemes \cite{rawlings2017model, sutton2018reinforcement}, for which we provide a high-level description in this section. To assess the efficiency of policies accounting for the peak power costs to minimise the sum payable of global REC bills over time, we introduce simpler variants which ignore the costs associated to the peaks when choosing the action. All the mathematical details about these policies are available in Appendix \ref{math_policies}.


\subsection{Optimal policies}

In this paper, we call admissible policies for an REC policies that provide actions that are feasible in the REC dynamical system. Formally, let $\Pi$ be the set of such policies:

\begin{align*}
\Pi = \Big\{ &\pi : S \times { \mathcal{H}_\mathcal{E}} \rightarrow \mathcal{U} \;  | \; \pi(s, e_{0:t})  \in U(s,e_{t}), \forall (s,e_{0:t}) \in \mathcal{S} \times { \mathcal{H}_\mathcal{E}}  \Big\}.
\end{align*} 

\noindent where $\mathcal{H}_\mathcal{E}$ is the set of all possible ordered sequences of exogenous variables defined as 

\begin{equation}
    \mathcal{H}_{\mathcal{E}} = \bigcup_{n \in \mathbb{N}^+}{\mathcal{E}^n}.
\end{equation}

\noindent We measure the performance of a policy when executed starting from a given state and a given sequence of past exogenous variable as the opposite of the expected discounted sum of the future costs arising from the actions chosen by the policy through time \cite{sutton1999policy}:

\begin{equation}
\label{eq:q_pi_value}
V_\pi(s_k, e_{0:k}) = \underset{T \rightarrow \infty}{\lim} \underset{\substack{e_{t+1} \sim P^{\mathcal{E}}(\cdot | e_{0:t}) }}{\mathop{\mathbb{E}}}{\sum_{t=k}^{T-1}{-\gamma^{t} \rho(s_t, e_t, u_t, s_{t+1})}}, 
\end{equation} where $\gamma \in ]0, 1[$ is a discount factor that indicates the relative importance of future costs compared to present costs. For any $t \geq k$, the next states are computed as $s_{t+1} = f(s_t, e_t, u_t)$ and the related action as $u_{t} = \pi(s_{t}, e_{0:t})$. We define the expected return of a policy $\pi \in \Pi$, starting from the initial state $S_0$, as follows:

\begin{equation}
\label{eq:objective_J_value}
J(\pi) = \underset{\substack{e_0 \sim P^{\mathcal{E}}_0(\cdot)}}{\mathop{\mathbb{E}}}{V_\pi(S_0, e_{0:0})}.
\end{equation}

\noindent The \emph{optimal REC control problem} is to identify a policy $\pi^* \in \Pi$ that maximises Equation \eqref{eq:objective_J_value}:
\begin{equation}
\label{eq:objective_pi}
\pi^* \in  \underset{\pi \in \Pi}{\arg \max} {\; J(\pi)}.
\end{equation}

\subsection{Model Predictive Control (MPC) policies}
\label{mpcpolicy}

\begin{figure}
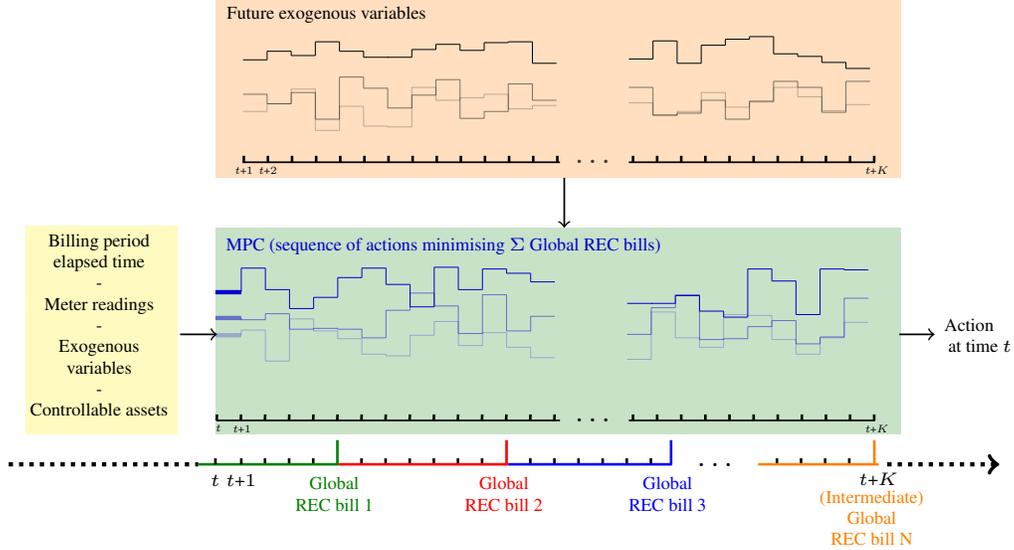

\centering
  \includestandalone[scale=1.145]{mpc_figure_V2}
  \caption{Illustration of how MPC policies compute the next action to apply in the REC dynamical system from a given state and a sequence of exogenous variables. The latter is composed of the current exogenous variable and values for the future ones (up to the policy horizon $t+K$). Colors are used to differentiate the billing periods that the MPC policies consider during its optimisation process.}
  \label{fig:mpcpolicy}
\end{figure}

The \emph{MPC policy} solves an internal optimisation problem minimising a discounted sum of the future costs (possibly over several billing periods) starting from a state and an exogenous variable sequence, and arising from a sequence of admissible actions up to a fixed time horizon $K$ that we denote as \emph{policy foresight}. This optimisation problem expects, as inputs, a sequence of future exogenous variables; they are usually computed with external methods, which are left out of the scope of this paper. If the last time step of this optimisation problem does not coincide with the end of a billing period, no cost signal is associated to the actions applied during this period. In that case, the objective function does not quantify the impact of these actions in the associated global REC bill. To avoid this pitfall, an intermediate global REC bill accounting for the elapsed time of that billing period, as defined in Section \ref{opt_realloc_scheme}, is added to the objective function at the last time step. To that end, we introduce a cost signal, namely $\widehat{GRB}$, which outputs this intermediate bill for time step $t+K$ if it is not located at the end of a billing period, and otherwise $0$. In short, the MPC policy computes the next (admissible) action $u^*_t$ at the time step $t$ such that
\begingroup
    \small%
\begin{equation}
    \widehat{GRB}_t = \argmax_{u_t} \left[ \max_{(u_{t'+1}, \ldots, u_{t'+K})} - \left(\widehat{GRB} + \sum_{t'=t}^{t+K}{\gamma^{t'} \rho(s_{t'}, \hat{e}_{t'}, u_{t'}, s_{t'+1})} \right) \right],
\end{equation}
\endgroup

\noindent where $\hat{e}_t = e_t$ is the current exogenous variable and $\hat{e}_{t+1:t+K}$ are future values of exogenous variables up to the policy foresight. We model this optimisation problem as a mixed-integer linear program (MILP). We also consider a variant of this policy, namely the \emph{MPC retail} policy, which only differs from the MPC policy in that the objective function of the MILPs does not include the costs related to the peaks. We have used CPLEX to solve the MILPs built by the MPC policies. Figure \ref{fig:mpcpolicy} provides an illustration of the computation steps of MPC policies.

\subsection{Reinforcement Learning (RL) policy}
\label{rlpolicy}

In Reinforcement Learning, one typically considers differentiable policies parameterised by a vector of parameters $\theta \in \Theta$. The goal is thus to identify the optimal parameters to maximize the expected return of the resulting policy:

\begin{equation}
\theta^* \in \argmax_{\theta \in \Theta}{J(\pi_{\theta})}.
\end{equation}

\noindent However, this policy cannot be computed directly as we do not possess the distributions $P_0^{\mathcal{E}}$ and $P^{\mathcal{E}}$. Instead, we focus on reinforcement learning algorithms performing stochastic gradient ascent on the objective function $J$, where the gradient is estimated through simulations of the policies in the POMDP. In particular, we use the Proximal Policy Optimization \cite{ppo} (PPO) algorithm. This algorithm learns, simultaneously with the parameterised policy, another parametric function, called the \emph{critic}. The latter is used, in combination with heuristic rules defined in the PPO algorithm, to mitigate the high variance issues associated with simulation-based gradient estimates. Neural networks are used as parametric functions, along with recurrent layers to consume sequences of exogenous variable values \cite{yu2019review, williams1990efficient}.

Policy gradient algorithms are often more efficient when costs are defined at every time steps. However, in the cost function defined in Section \ref{probstat}, sparsity increases with the duration of a billing period (the longer the billing period gets, the more we have zeros in the costs). To attempt to mitigate this issue, we consider another version of this policy, denoted \emph{RL dense}, in which zero costs are replaced by intermediate global REC bills during the policy search.

Similarly to the policy \emph{MPC retail}, we consider a variant of this policy that we denote as \emph{RL retail}. In this variant, cost signals are modified, before updating the parameters of the policy, as if there were no power peak costs (i.e., assuming that $P^- = P^+ = 0$). We also define a policy which combines the cost modifications applied in the policy search procedure for the RL dense and RL retail policies. We refer to that policy as \emph{RL dense retail}. To implement the RL policies with the above-mentioned procedure, we have used the PPO implementation of the Python programming library \emph{RLlib} \cite{rllib} to implement this policy search procedure, along with the deep learning Python library \emph{PyTorch} \cite{pytorch} to implement the neural network models.

\section{Experiments}
\label{testpolicies}
In this section, we simulate the policies described in Section \ref{policies}. Note that, in practice, we do not simulate algorithms for an infinite time; in our experiments, we always choose a time horizon that is sufficiently long to cover several billing (and consequently, market) periods. Due to the scarcity of historical data associated with existing RECs, we sample exogenous variables (up to the fixed time horizon of the simulations) by applying a time-correlated white noise to (scarcely) existing data; see Appendix \ref{sample_exo_algo} for more details about this sampling procedure. We first describe how future exogenous values are computed for the MPC policies in the context of our simulations. We then introduce baseline policies in Section \ref{baselinepol} that we compare with MPC and RL policies against two REC instances, referred to as \emph{REC-2} and \emph{REC-7}; they respectively count two and seven members, see Appendices \ref{rec_2_details} and \ref{rec_7_details} for more details about the experimental protocol settings. The first one is an illustrative example built from synthetic data, and the second one from historical data from an existing REC located in Wallonia, Belgium. We compare the results of all the simulated policies, respectively for each REC instance, in Sections \ref{rec_2_results} and \ref{rec_7_results}. The settings related to the procedure to build the RL policies follow the usual recommendations related to policy gradient algorithms \cite{whatmattersonpolicy, recurrentppo}, as detailed in Appendix \ref{rlpoliciesdetails}. 

\subsection{Computing future exogenous values for MPC policies}
\label{mpc_policy_predict}
MPC policies expect future values of exogenous variables to compute its next action. In a realistic setting, these future values, usually computed with forecasting methods \cite{cons_forecast, prod_forecast}, differ from the actual future values; this difference often increases with the length of these values, and has an (often negative) impact in the performance of the MPC policies. As mentioned earlier in this section, sequences of exogenous variables are sampled, from existing data, at once before simulating the policies. We thus propose the following procedure to compute the future values of the exogenous variables (during the simulation of the MPC policies). At each time step $t$, the future exogenous value at the next time step $t+1$, that we provide to the MPC policies, corresponds to the actual one. From $t+2$ (if $K > 1$), the time series of future exogenous values progressively converges (from the true future values) to the existing data (again, from $t+2$). We characterise the speed of this convergence by a parameter $\alpha \in ]0, 1]$, such that if $\alpha=0$, then the time series provided to the MPC policies is equals to the existing data from $t+2$; and that if $\alpha = 1$, then all the next exogenous variables are equal to the true future values (again, from $t+2$). In this procedure, another characteristic is that, as long as $\alpha$ gets closer to $0$, the convergence speed of the predicted time series (to the existing data) dramatically increases; see Appendix \ref{mpc_predicting_algo} for a more detailed description of this sampling procedure. We refer to this parameter ($\alpha$) as the \emph{foresight efficiency}.

\subsection{Baseline policies}
\label{baselinepol}

We introduce the following baseline policies to provide boundaries in which the expected returns of the MPC and RL policies are located:

\paragraph{Self-consumption.}

The two following rule-based policies output the next actions to be applied to the controllable assets by maximising some self-consumption criteria. The first one, denoted as the \emph{REC policy}, compares the total net consumption and the total net production of the non-controllable assets of the REC. It then uses the controllable assets, to either absorb the excess of electricity consumption or the surplus of electricity production. The second one, denoted as the \emph{SELF policy}, uses the same approach but individually for each member equipped with controllable assets.

\paragraph{Perfect-foresight.}

We consider the particular case of MPC policies where $\alpha = 1$ and $K = T$ with $T$ a finite time horizon that is fixed for the simulations. In such a case, as defined in Section \ref{mpcpolicy}, the MPC policies perfectly "predict" the future exogenous values at each time step until the time horizon $T$, thus leading to the minimal discounted sum of the global REC bills for that horizon. Under these conditions, the MPC retail policy leads to the minimal discounted sum of the global REC bills with $P^+ = P^- = 0$. We refer to these respective policies as the \emph{OPT policy} and the \emph{OPT retail}.


\subsection{REC-2}
\label{rec_2_results}
REC-2 is composed of two members. One of them is only equipped with non-controllable assets that consume electricity; the other one is equipped with PV panels, that always produce more electricity than is consumed by the other non-controllable assets, and is also equipped with a battery. Their corresponding consumption and production profiles (from non-controllable assets) are derived from synthetic historical data; see Appendix \ref{rec_2_details} for more details about these profiles. We set the time horizon of the simulations at $T=101$. Expected returns of baseline policies and MPC policies are estimated over $1024$ simulations. In order to evaluate the RL policies, we first train $16$ policy functions using different random seeds. This training procedure is carried as follows. Each policy function is trained for $600$ iterations with the PPO algorithm, each iteration being comprised of $64$ simulations of the policy functions. At the end of these simulations, we have $16$ instances of trained RL policies. We estimate their expected returns by running $64$ simulations for each of these RL policies, and we average them.

Figure \ref{fig:rec2results} shows the expected returns of all the simulated policies in REC-2; see Appendix \ref{rec_2_details} for more details about the simulation settings of REC-2. We first notice that the expected returns of MPC policies quickly converge as $K$ grows. Furthermore, MPC policies with perfect foresight efficiency ($\alpha = 1$) yield optimal expected returns (with respect to the sampled exogenous variables time series). We observe that the SELF policy has the worst expected return. Since the owner of the battery only produces electricity from non-controllable assets, this policy will only charge the battery until full capacity is reached. The profiles show that no electricity production is available for the first time steps. As a consequence, the producer turned to consume electricity from the main network by charging its battery, thus increasing the size of energy consumption bills as well as the offtake peaks, at least for the first billing period. The REC policy mitigates this issue by accounting for the electricity consumption of all the members. However, the other policies (eventually) yield better expected returns than these rule-based policies. The performance of RL policies ignoring peak costs are close to REC policy; they might have learnt to operate the battery similarly to the latter. Although the RL policies have a worse expected return than the OPT retail policy, it manages to perform slightly better than some of the MPC policies with noisy prediction. This illustrates the advantage of having access to predictions of future exogenous variable values (given that the foresight efficiency is reasonably high). Note that the expected return of RL dense is slightly worse than the RL policy. 

During the simulations of the PPO algorithm (see Appendix \ref{rec_2_details} for detailed results), we also noticed that the RL dense policy got a slightly better expected return, for the first iterations, than the RL policy (we have even observed a higher gap for RL retail and RL retail dense). Both observations are due to the modified cost signal (by adding intermediate global REC bills, with or without peak costs). Indeed, having cost signals at every time steps allowed to speed up the policy updates (in terms of expected return) for the first few iterations. However, the discounted sum of the original cost signals is not equal to the discounted sum of the modified cost signals. As a consequence, the RL dense policy might have converged too quickly to a local optimum due to the difference between the two formulations.

\begin{figure}
\centering
\includegraphics[width=13.7cm]{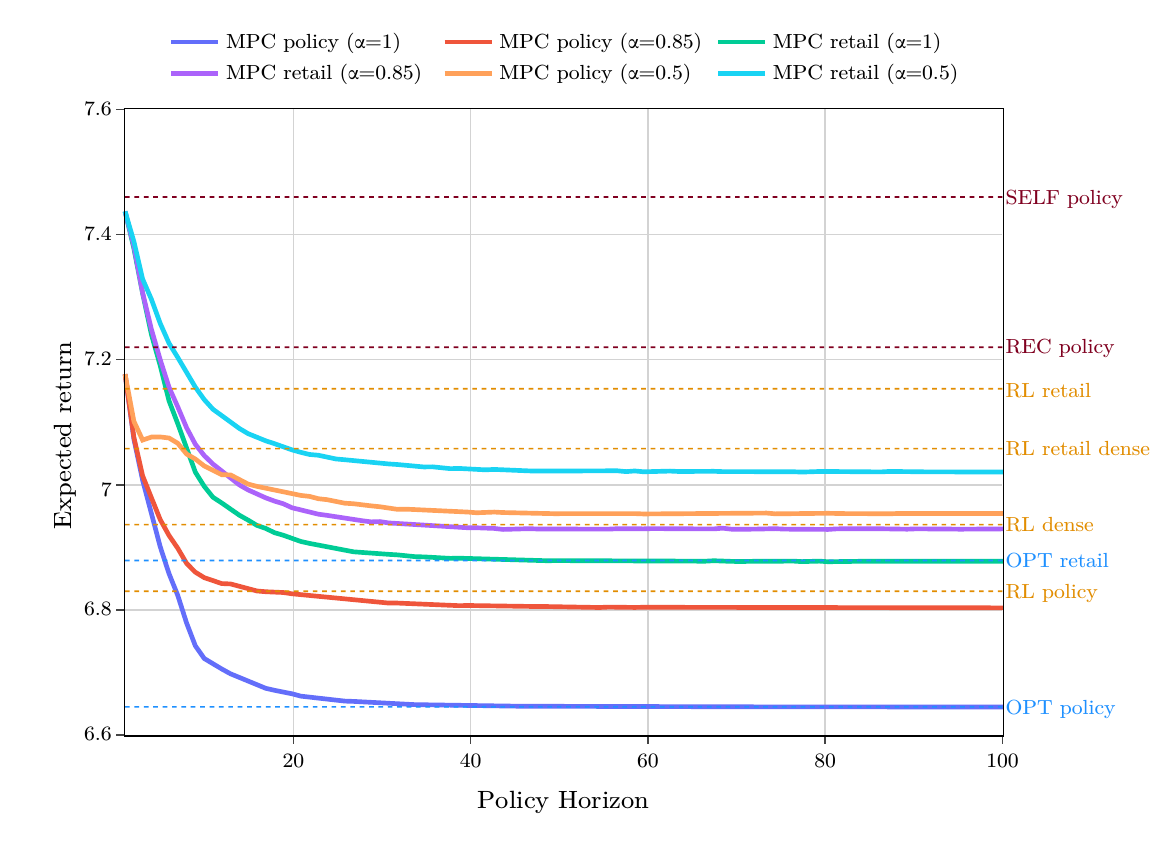}
\vspace{-0.33cm}
\caption{Expected returns of MPC policies (averaged over $16$ runs) in REC-2, given policy horizon and foresight efficiency, along with expected returns of RL and baselines policies. Recall that $\alpha$ is the foresight efficiency defined in Section \ref{mpcpolicy}, with $\alpha=1$ corresponding to perfect foresight.}
\label{fig:rec2results}
\end{figure}

\subsection{REC-7}
\label{rec_7_results}

REC-7 is composed of seven members. Some of them are equipped with PV panels. Similarly to REC-2, one of them is equipped with a (controllable) battery. Consumption and production profiles are derived from available historical data from an existing REC located in Wallonia, Belgium; see Appendix \ref{rec_7_details} for more details. We set the time horizon of the simulations at $T=721$. Expected returns of baseline policies and MPC policies (with their respective policy horizons and foresight efficiencies) are estimated at $4096$ simulations. Expected returns of baseline policies and MPC policies are estimated over $1024$ simulations. We evaluate the RL policies in an identical manner to those for REC-2, excepted that we run the PPO algorithm on $32$ policies for $1000$ iterations and that we perform, for each policy, $128$ simulations at each iteration to train them, and another $128$ simulations to estimate the expected return of the RL policies.

\begin{figure}
\centering
\includegraphics[width=13.5cm]{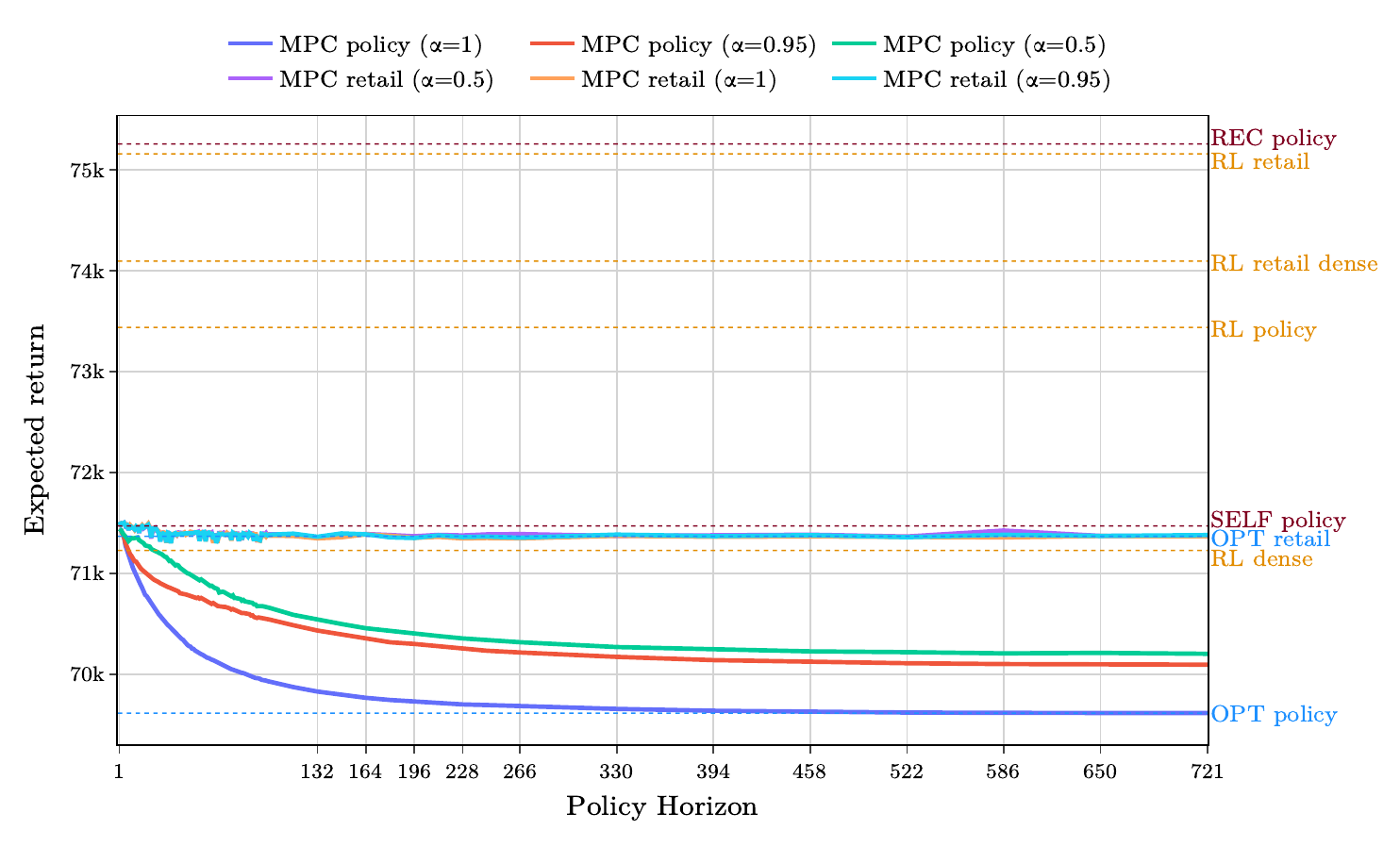}
\vspace{-0.33cm}
\caption{Expected returns of MPC policies (averaged over $16$ runs) in REC-7, given policy horizon and foresight efficiency, along with expected returns of RL and baselines policies.}
\label{fig:rec7results}
\end{figure}

Similarly to Section \ref{rec_2_results}, we report the results for simulating the policies in REC-7 in Figure \ref{fig:rec7results}. We first notice that REC policy has a worse expected return than the SELF policy. This is surprising, and might be due to the composition of this REC. More surprisingly, RL policies ignoring peak costs also have a worse expected return than the SELF policy. The RL policy barely managed to achieve better results than the latter. They might have suffered from either the high sparsity of the cost function, or the dense formulation (with intermediate global REC bills) of the discounted sum of costs (for the RL retail dense). Only RL dense managed to (slightly) achieve better than the OPT retail policy. While MPC policies are still the best policies in REC-7 (beating RL dense with a rather low policy foresight), their variants ignoring the peak costs are still better than most of the policies.

\subsection{General comments}

Overall, MPC policies and RL policies are better than self-consumption rule-based policies, excepted for REC-7 where the SELF policy is better than some RL policies, but notably not the RL dense one. Note that, with the exception of the REC policy in REC-7, the expected returns of the policies were not very far from each other. This might be due to the limited impact of the controllable assets in the cost savings \cite{FELICE2022119419}. As expected, MPC policies accounting for the peak costs achieved the best-expected returns with a reasonably low policy foresight. However, as shown in the results, they are dependent on both the policy horizon and the foresight efficiency. In our results, the expected returns of RL policies were, at best, close to the OPT retail policy.

Both reshaping the reward signals and accounting for the peak costs prove to be useful. Indeed, in the case of REC-7, RL dense obtains better results than the other RL policies. However, ignoring the peak costs in the reward signals has a limited effect (RL retail dense) compared to simple baseline policies, as well as taking into account the peak costs without densifying the reward signals (RL policy). This is not true for REC-2 where the non-densified RL policy (accounting for the peak costs) beats RL dense. This is probably mainly due to the limited time horizon, providing an already relatively dense reward signal.

The required computation time for computing the next action with MPC policies dramatically increases with the number of members and the policy horizon, as shown in Table \ref{tab:computation_times}. Conversely, and similarly to the baseline policies, computing the next action with RL policies (once trained) is much faster and scales with large sizes of RECs. However, the overall runtime of training RL policies may dramatically increase (with the size of RECs), especially with the frequency of the reward function computation (e.g., at every time steps).

There is thus a trade-off between MPC policies (very good results but slow) and RL policies (not as good results but very fast at runtime while very slow to train) that must be made depending of the REC-specific needs.

\begin{table}
        \centering
\begin{tabular}{c rrrr rr rr}
    \toprule
  & \multicolumn{4}{c}{\shortstack{MPC policies}} & \multicolumn{2}{c}{\shortstack{RL policies}} & \multicolumn{2}{c}{\shortstack{Baselines}} \\
    \cmidrule(lr){2-5} \cmidrule(lr){6-7}  \cmidrule(lr){8-9}   
       REC instance & $\mathrm{\frac{T}{8}}$ & $\mathrm{\frac{T}{4}}$ & $\mathrm{\frac{T}{2}}$ & T & Update & Action & REC & SELF     \\ 
        \cmidrule(lr){0-0} \cmidrule(lr){2-5} \cmidrule(lr){6-7}  \cmidrule(lr){8-9}   
        REC-2 & $\mathbf{0.01}$ & $\mathbf{0.02}$ & $\mathbf{0.04}$ & $\mathbf{0.82}$ & $\mathbf{43.20}$ & $\mathbf{6\mathrm{\mathbf{e}}{-4}}$ & \multicolumn{2}{c}{\shortstack{$\mathbf{1\mathrm{\mathbf{e}}{-3}}$}}\\
        \cmidrule(lr){0-0} \cmidrule(lr){2-5} \cmidrule(lr){6-7}  \cmidrule(lr){8-9}     
        REC-7 & $\mathbf{0.05}$ & $\mathbf{0.11}$ & $\mathbf{0.20}$ & $\mathbf{0.40}$ & $\mathbf{321.27}$ & $\mathbf{1\mathrm{\mathbf{e}}{-4}}$  & \multicolumn{2}{c}{\shortstack{$\mathbf{1\mathrm{\mathbf{e}}{-3}}$}} \\

    \bottomrule
\end{tabular}
        \vspace{0.33cm}
        \caption{Sample computation times required to compute the next action for MPC policies (given policy horizons), RL policies (also including update iteration of the PPO algorithm) and baselines (SELF and REC) policies. Times are expressed in seconds, and are averaged over $32$ independent runs. These computation times have been performed on a laptop equipped with $32$ GB of RAM and a (Intel Gen Core i7) CPU with $12$ cores and $16$ threads.}
        \label{tab:computation_times}
     \end{table}

\section{Conclusion and Future work}
\label{conclusion}

In this paper, we have proposed a generic modelling framework to optimise RECs with controllable assets towards the minimisation of the sum payable of the \emph{ex-post} electricity bills (accounting for the energy exchanges through the REC), which we call the global REC bill. Prominent elements in these electricity bills are the network peak fees, which are computed from the energy exchanges with the retailers. In this framework, we have formalised how the Energy Community Manager optimally reallocates (to minimise the sum payable of the ex-post electricity bills), at regular time intervals, the surplus of electricity production to the members. This formalisation is integrated in the cost function of a Partially Observable Markov Decision Process, which also encapsulates the dynamics related to the electricity power flows (accounting for the controllable assets). We have tested practical policies that approximate the optimal ones, from the model predictive control and reinforcement learning classes of techniques. These policies have been compared to baseline algorithms, including variants of these policies that ignore the peak costs, or rule-based policies that operate the controllable assets to maximise self-consumption criteria. The tests were carried against an artificial REC instance, and against an REC instance derived from historical data of an existing REC located in Wallonia, Belgium. The results obtained from these tests show how policies, accounting for the peak costs, may sensitively decrease the global REC bills. They also strongly encourage pursuing the development of REC frameworks to optimise the controllable assets from a centralised standpoint, possibly by addressing the following limitations of the approach proposed in this paper.

The optimal reallocation scheme, described in Section \ref{opt_realloc_scheme}, is computed by solving an optimisation problem, which is done in practice through third-party solvers. As long as the size of the RECs stays relatively low, solving this reallocation scheme is quite fast. But as RECs grow, with dozens, even hundreds of members, commercial third-party solvers coupled with heavy-duty hardware begin to be required to scale the computation time, especially when simulations are needed for e.g., running business cases for REC investments, or training algorithms (like RL policies) to optimally operate RECs with controllable assets. However, these solvers usually exploit interior point methods to solve convex optimisation problems. Running these algorithms with GPUs may be much faster compared to non-commercial third-party solvers \cite{agrawal2019differentiable}\cite{optlayer}\cite{wu2017scalable}. Another research direction would be to pursue the efforts to identify closed-form approaches to compute approximations for optimal reallocation schemes (accounting for the peak costs) which still lead to efficient policies that rely on this approximation.

Similar issues arise for MPC policies. Indeed, they require one to solve mixed-integer linear programs (see Appendix \ref{math_mpc_policies} for more details). As the size of the REC and the policy horizon both grow, their computation time dramatically increases. These policies also require predictions of the future electricity flows of the non-controllable assets of the REC members. Scarcity of historical data, controllable assets with complex dynamics, and the difficulty to build algorithms dedicated to predictions of these flows make it challenging to scale MPC techniques with large RECs. Similar to the optimal reallocation scheme, adapting these MPC policies to run on GPUs might be an alternative to solve MILPs [ref needed]. As for the electricity flows predictions, to make up for the scarcity of historical data for a given REC, transfer learning techniques might help, provided supervised learning models predicting similar electricity flows exist \cite{sarmas2022transfer}.

The RL policies have been shown to be a promising alternative to MPC policies, especially when the reward signals are densified, as they keep a rather low complexity to be used to operate RECs with controllable assets once they have been trained. For the training procedure, since the REC control problem is centralised (by the ECM), we have used a single-agent policy gradient algorithm called PPO \cite{ppo}. While building RL policies through this single-agent configuration should be the proper approach, they are in practice difficult to train due to both the presence of recurrent layers \cite{recurrentppo} and the amount of information needed to feed to the parametric functions used by the RL policies, which quadratically grows with the number of members in the REC and the length of a billing period. The complexity of the cost signal, which is mainly due to the peak costs and which requires the solving of an optimisation problem, continue to make the scaling of the training procedure for large RECs difficult. Yet, according to our results, the presence of these peak costs have significantly impacted the expected return of these RL policies, particularly in REC-7, our largest REC in our simulations having seven members. To mitigate these scaling issues, a surrogate cooperative multi-agent POMDP, where each member could be an agent, might be designed to build the RL policies, while maintaining their compatibility with the original (centralised) POMDP \cite{yu2022surprising, zeroenergymultiagent, marlpeakrebounds}.

\bibliographystyle{unsrt}
\bibliography{main}
\newpage

\appendix
\section{Mathematical details}

\subsection{Decision process}
\label{mdpmath}

In this section, we provide a detailed description of the decision process introduced in Section \ref{probstat} as a particular instance of a Partially Observable Markov Decision Process (POMDPs). This POMDP provides the following time discretisation. The duration between two consecutive time steps (for controllable assets and thus meter reading transitions) is defined by the value $\delta \in \mathbb{R}^+$. The duration of a market period, expressed in the number of time steps, is defined by the value $\Delta_M \in \mathbb{N}^+$. The duration of a billing period, expressed in number of market periods, is defined by the value $\Delta_B \in \mathbb{N}^+$. We describe below, in detail, the components of the POMDP. Let $M$ be the number of members of the REC. A state $s \in \mathcal{S}$ contains the following information:

\begin{itemize}
    \item the number of time steps elapsed in the current market period $s^{\Delta_M} \in \mathbb{N}$,
    \item the number of market periods elapsed in the current billing period $s^{\Delta_B} \in \mathbb{N}$,
    \item the state of the controllable assets $s^c_m$ of each member $m \in \left\{1, \ldots, M\right\}$,
    \item the production meter reading $s^{r_+}_{(m, n)} \geqslant 0$ of each member $m \in \left\{1, \ldots, M\right\}$ at each market period $n \in \left\{1, \ldots, \Delta_B\right\}$,
    \item the consumption meter reading $s^{r_-}_{(m, n)} \geqslant 0$ of each member $m \in \left\{1, \ldots, M\right\}$ at each market period $n \in \left\{1, \ldots, \Delta_B\right\}$.
\end{itemize}

\noindent An exogenous variable $e \in \mathcal{E}$ contains the following information:
\begin{itemize}
    \item the electricity net production of each member $e_m^p \geqslant 0$,
    \item the electricity net consumption of each member $e_m^c \geqslant 0$.
\end{itemize}

\noindent Since these net electricity flows account for self-consumption during the time interval between $t$ and $t+1$, we have that $e_m^p e_m^c = 0$. The dynamics of the controllable assets are derived from their specifications. The dynamics related to the states memorising the elapsed time in the current billing period are defined as follows:

\begin{align}
\label{eq:detailed_elapsed_time}
s_{0}^{\Delta_M} &= s_{0}^{\Delta_B} = 0,\\
s_{t+1}^{\Delta_M} &= \begin{cases}
    1,& \text{if } s_{t}^{\Delta_M} = \Delta_M,\\
     s_{t}^{\Delta_M} + 1,              & \text{otherwise,}
\end{cases}\\
s_{t+1}^{\Delta_B} &= \begin{cases}
    0,& \text{if } s_{t}^{\Delta_B} = \Delta_B,\\
     s_{t}^{\Delta_B},              & \text{if } s_{t}^{\Delta_M} < \Delta_M,\\
     s_{t}^{\Delta_B}+1,              & \text{otherwise. }\\
\end{cases}
\end{align}

\noindent Let $q_m^c$ be a function giving the amount of electricity consumption (positive) or production (negative) generated by applying the action $u_m^c$ in the controllable asset state for each member $s_m^c$. The dynamics related to the production and consumption meter readings are defined for all members $m \in \left\{1, \ldots, M \right\}$  as follows:

\begin{align}
\label{eq:detailed_meters_dyn}
s_{(m, n), 0}^{r_+} &= s_{(m, n), 0}^{r_-} = 0, \; \forall n \in \left\{ 1, \ldots, \Delta_B\right\} \\[0.3cm]
s_{(m, n), t+1}^{r_+} &= \begin{cases}
    0 & \text{if } s_{t}^{\Delta_B} = \Delta_B,\\
     s_{(m, n), t+1}^{r_+}             & \text{if $s_{t}^{\Delta_B} > n-1$,}\\
     s_{(m, n), t+1}^{r_+} +  l_{m, t}^+           & \text{otherwise,}\\
\end{cases}\\[0.3cm]
s_{(m, n), t+1}^{r_-} &= \begin{cases}
    0 & \text{if } s_{t}^{\Delta_B} = \Delta_B,\\
     s_{(m, n), t+1}^{r_-}             & \text{if $s_{t}^{\Delta_B} > n-1$,}\\
     s_{(m, n), t+1}^{r_-} +  l_{m, t}^-           & \text{otherwise,}\\
\end{cases}
\end{align}
\noindent where 
\begin{align}
l^+_{m, t} &= \max(q^c_m(s_{m, t}^c, u_{m, t}^c) - e^c_{m, t} + e^p_{m, t}, 0.0),\\
l^-_{m, t} &= \max(q^c_m(s_{m, t}^c, u_{m, t}^c) + e^c_{m, t} - e^p_{m, t}, 0.0).
\end{align} are respectively the net electricity production (i.e., after accounting for self-consumption) and the net electricity consumption (i.e., after accounting of self-consumption) of the member $m$ at time step $t$.

As defined in Section \ref{pomdp_descr}, the cost signal of the POMDP is computed by solving the minimisation problem defined in Equation \eqref{eq:objreallocaoptim}, resulting in what we call the global REC bill, from input values (related to meter readings and elapsed time in the billing period) provided by states at the last time steps of any billing periods; otherwise the cost signal is zero. However, intermediate global REC bills can be computed during the simulation of the POMDP by adapting the value of $\tau$, introduced in Equation \eqref{eq:objrealloc}, to correspond to the time elapsed during the billing period. More formally, given decision variables $e^+_m$ and $e^-_m$ for each member $m \in \left\{1, \ldots, M \right\}$, related to the energy exchanged by the members through the REC as defined in Section \ref{opt_realloc_scheme}, we rewrite Equation \eqref{eq:objrealloc} to account for the input values provided by a state $s \in \mathcal{S}$ as follows:

\begingroup
\small
\begin{align}
\label{eq:mathobjreallocpomdp}
\hspace{-2cm} \nonumber &EB''_m(e^+_m, e^-_m, s) = \\[0.5em] &\left( \sum_{n=1}^{\delta_B}{B_{m}^- (s_{(m, n)}^{r^-} - e_{m, n}^-) - B_{m}^+ (s_{(m, n)}^{r^+} - e_{m, n}^+) + \Lambda^- e_{m, n}^- + \Lambda^+ e_{m, n}^+} \right) \\[0.5em]
\nonumber & +\frac{s^{\Delta_{BM}}}{\Delta_B\Delta_M} \left[ P^- \left(\max_{n \in \left\{1, \ldots, \delta_B\right\}}{s_{(m, n)}^{r^-} - e_{m, n}^-}\right) + P^+ \left( \max_{n \in \left\{1, \ldots, \delta_B\right\}}{s_{(m, n)}^{r^+} - e_{m, n}^+} \right) \right],
\end{align} \endgroup where $\delta_B = \min(s^{\Delta_B}+1, \Delta_B)$, and $s^{\Delta_{BM}}$ is the number of time steps elapsed during the billing period corresponding to the state $s$:

\begin{equation}
s^{\Delta_{BM}} = \begin{cases}
    \Delta_M\Delta_B & \text{if } s^{\Delta_B} = \Delta_B,\\
     s^{\Delta_B} \Delta_M             & \text{if } s^{\Delta_M} = \Delta_M,\\
     s^{\Delta_B} \Delta_M + s^{\Delta_M}             & \text{otherwise.}\\
\end{cases}
\end{equation}

\noindent 

\subsection{Policies}
\label{math_policies}

In this section, we provide a detailed description of the policies introduced in Sections \ref{policies} and \ref{baselinepol} in the context of the POMDP described above. Namely, (i) the MPC policies which decide, given an approximate prediction of the future exogenous values, the next optimal action to apply (with respect to the time horizon and the values of the prediction); (ii) the RL policies which approximate the optimal policies through parameterised functions that output the next action by exploiting only the current state and history of exogenous values; (iii) the SELF policy and the REC policies that outputs the next action which maximises some self-consumption criteria given the current state and exogenous variable values.

\subsection{MPC Policies}
\label{math_mpc_policies}
The MPC policies compute the next action by solving an optimisation problem accounting for the structure of the POMDP introduced in Section \ref{pomdp_descr} and future exogenous variable values; we refer to this sequence of values as $\hat{e}$. More formally, these policies compute the optimal sequence of actions that minimises, with respect to the current state and the predicted exogenous values, the discounted sum of the upcoming global REC bills. To that end, it solves a mixed-integer linear (MILP) program that is modelled as follows. Let $\mathcal{T}^{\Delta_M}_{t: t+K}$ be the last time steps of market periods between $t$ and $t+K$:
\begin{equation}
\mathcal{T}^{\Delta_M}_{t: t+K} = \left\{ t' \in \left\{ t-s_{t'}^{\Delta_{BM}}, \ldots, t+K \right\} | s_{t'}^{\Delta_M} = \Delta_M \; \textbf{or} \; t' = t+K \right\}.
\end{equation}
\noindent Let $(g^-_{m, t^{\Delta_M}}, g^+_{m, t^{\Delta_M}}, r^-_{m, t^{\Delta_M}}, r^+_{m, t^{\Delta_M}})$ be the non negative auxiliary variables that respectively correspond, for each member $m \in \left\{1, \ldots, M \right\}$ and for each last time step market period $t^{\Delta_M} \in \mathcal{T}^{\Delta_M}_{t: t+K}$, to the amount of energy bought(sold) from(to) the retailer, as well as energy bought(sold) from(to) the REC (local market). Let $\mathcal{T}^{\Delta_B}_{t+1: t+K}$ be the time steps that correspond to the last time step of billing periods between $t+1$ and $t+K$:
\begin{equation}
\mathcal{T}^{\Delta_B}_{t+1: t+K} = \left\{ t' \in \left\{ t+1, \ldots, t+K \right\} | s_{t'}^{\Delta_B} = \Delta_B \;  \textbf{or} \; t' = t+K \right\}. 
\end{equation}
Let $p^-_{m, t^{\Delta_B}}$ and $p^+_{m, t^{\Delta_B}}$ be the non negative auxiliary variables that respectively correspond to the offtake and injection peaks for each member $m \in \left\{1, \ldots, M \right\}$ and for each end of billing period $t^{\Delta_B} \in \mathcal{T}^{\Delta_B}_{t: t+K}$. We concisely denote the energy exchanges summed over an entire billing period, which ends at $t^{\Delta_B} \in \mathcal{T}^{\Delta_B}_{t+1: t+K}$, as follows:
\begin{align}
g^-_{m, t^{\Delta_B}} &= \sum_{t^{\Delta_M} \in \mathcal{T}^{\Delta_{B \leftarrow M}}_{t^{\Delta_B}}}{g_{m, t^{\Delta_M}}^-}, \\ \nonumber
g^+_{m, t^{\Delta_B}} &= \sum_{t^{\Delta_M} \in \mathcal{T}^{\Delta_{B \leftarrow M}}_{t^{\Delta_B}}}{g_{m, t^{\Delta_M}}^+}, \\ \nonumber
r^-_{m, t^{\Delta_B}} &= \sum_{t^{\Delta_M} \in \mathcal{T}^{\Delta_{B \leftarrow M}}_{t^{\Delta_B}}}{r_{m, t^{\Delta_M}}^-}, \\ \nonumber
r^+_{m, t^{\Delta_B}} &= \sum_{t^{\Delta_M} \in \mathcal{T}^{\Delta_{B \leftarrow M}}_{t^{\Delta_B}}}{r_{m, t^{\Delta_M}}^+}, \\ \nonumber
\end{align}

\noindent where $\mathcal{T}^{\Delta_{B \leftarrow M}}_{t^{\Delta_B}} = \mathcal{T}^{\Delta_M}_{t^{\Delta_B} - s_{t'}^{\Delta_{BM}}: t^{\Delta_B}}$ is the set of time steps corresponding to end of market periods that belong to the billing period ending at time step $t^{\Delta_B}$. To compute the next action, the MPC policies select an optimal sequence (of decisions) minimising the following objective function (discounted sum of the upcoming global REC bills):

\begingroup
\footnotesize
\begin{align}
&\widehat{GRB}(s_t, \hat{e}_{t:t+K}, \hat{u}_{t:t+K}) = \\ \nonumber &\underset{\left(g^-_{m, t^{\Delta_M}}, g^+_{m, t^{\Delta_M}}, r^-_{m, t^{\Delta_M}}, r^+_{m, t^{\Delta_M}}\right), \; \forall t^{\Delta_M} \in \mathcal{T}^{\Delta_B}_{t: t+K}}{\min} \sum_{t^{\Delta_B} \in \mathcal{T}^{\Delta_B}_{t: t+K}}
   \gamma^{(t^{\Delta_B} - t)} \sum^M_{m=1} \\ \nonumber &\Bigg[ B_{m}^- g_{m, t^{\Delta_B}}^- - B_{m}^+ g_{m, t^{\Delta_B}}^+ + \Lambda^- r_{m, t^{\Delta_B}}^- + \Lambda^+ r_{m, t^{\Delta_B}}^+  + \frac{P^+ p^+_{m, t^{\Delta_B}} + P^- p^-_{m, t^{\Delta_B}}}{s_{t^{\Delta_B}}^{\Delta_{BM}}} \Bigg]
\end{align}
\endgroup
\noindent where $\gamma$ is the discount factor associated to the POMDP. Note that MPC retail only differs from the MPC policy in that it sets $P^+ = P^- = 0$ in its internal optimisation problem. The decision variables of the MILP, which correspond to a sequence of actions in the POMDP, are constrained accordingly to the function $U$ defined in Section \ref{probstat}. Since these constraints depend on the current state (of the controllable assets), we introduce auxiliary variables for the sequence of states for each member $m$ equipped with controllable assets. In these sequences, the first state corresponds to $s_t$, while the subsequent states are defined through the transition function $f_m^c$. For each member, the variables related to the offtake and injection peaks correspond, for a given billing period ending at time step $t^{\Delta_B} \in \mathcal{T}^{\Delta_{B}}_{t^{\Delta_B}}$, to the maximum values of the energy exchanged with the retailer across the market periods within this billing period:

\begin{align}
p^-_{m, t^{\Delta_B}} \geqslant g^-_{m, t^{\Delta_M}},\\
p^+_{m, t^{\Delta_B}} \geqslant g^+_{m, t^{\Delta_M}},
\end{align}

\noindent for all $m \in \left\{1, \ldots, M\right\}$ and $t^{\Delta_M} \in \mathcal{T}^{\Delta_{B \leftarrow M}}_{t^{\Delta_B}}$. We introduce, in the optimisation problem modelled by the MPC policies, the constraints related to the energy exchanges described in Section \ref{opt_realloc_scheme} as follows: 

\begingroup
\small
\begin{align}
  g^-_{m, t^{\Delta_M}} +
  r^-_{m, t^{\Delta_M}}
 &= \begin{cases}
        s^{r, -}_{t} + l^-_{m, t:t^{\Delta_M}} & \text{if } t^{\Delta_M} > t \text{ and }t^{\Delta_M} - \Delta_M < t, \\
        l^-_{m, (t-t^{\Delta_M}):t^{\Delta_M}}& \text{if } t^{\Delta_M} > t, \\
        s^{r, -}_{t^{\Delta_M}} & \text{otherwise,}
    \end{cases}\\ \nonumber
  g^+_{m, t^{\Delta_M}}+
  r^+_{m, t^{\Delta_M}}
 &= \begin{cases}
        s^{r, +}_{t} + l^+_{m, t:t^{\Delta_M}} & \text{if } t^{\Delta_M} > t \text{ and }t^{\Delta_M} - \Delta_M < t, \\
        l^+_{m, (t-t^{\Delta_M}):t^{\Delta_M}}& \text{if } t^{\Delta_M} > t, \\
        s^{r, +}_{t^{\Delta_M}} & \text{otherwise,}
    \end{cases},\\ \nonumber
    \sum^{M}_{m=1} r^-_{m, t^{\Delta_M}} &= \sum^{M}_{m=1} r^+_{m, t^{\Delta_M}},
\end{align}
\endgroup

\noindent for all $t^{\Delta_M} \in \mathcal{T}^{\Delta_M}_{t: t+K}$, where
\begin{equation}
\label{eq:net_pmeters_mpc}
  l^-_{m, t_{min}:t_{max}} = \max\left(\sum_{t'=t_{min}}^{t_{max}}{l^-_{m, t'}} - \sum_{t'=t_{min}}^{t_{max}}{l^+_{m, t'}}, 0\right)
\end{equation}
\noindent and 
\begin{equation}
\label{eq:net_cmeters_mpc}
  l^+_{m, t_{min}:t_{max}} = \max\left(\sum_{t'=t_{min}}^{t_{max}}{l^+_{m, t'}} - \sum_{t'=t_{min}}^{t_{max}}{l^-_{m, t'}}, 0\right)
\end{equation}
\noindent are respectively the net electricity consumption and production from time step $t$ to $t' > t$. We introduce auxiliary variables to compute the net electricity flows (for members that are equipped with controllable assets). To account for self-consumption, we define the value of these variables as follows. If $l^- > l^+$, then the variable corresponding to the net consumption is equals to $l^+ - l^-$. Similarly, if $l^+ > l^-$, then the variable corresponding to the net production is equals to $l^- - l^+$. These two variables are mutually exclusive, only one of them can be non-zero at any time ($l^- l^+ = 0$). For net electricity flows involving controllable assets, we have implemented this mutual exclusion constraint by using \emph{special ordered sets} \cite{special_ordered_set}. We have followed the same approach to implement Equations \eqref{eq:net_pmeters_mpc} and \eqref{eq:net_cmeters_mpc}.

\subsection{RL Policies}
\label{math_rl_policies}

Unlike the MPC policies, the RL policies compute the next action without an explicit approximation of the future exogenous values by using parameterised closed-form functions which are differentiable (with respect to their parameters). In this section, we describe how this policy computes the next action through parameterised functions. Thereafter, we describe how these parameters are determined through reinforcement learning to approximate the optimal policies.

We consider stochastic and parameterised functions which are differentiable with respect to their parameters (usually, deep neural networks). These functions take as input the history of exogenous variables and the states, which are usually transformed (e.g., to decrease dimensionality), and output the parameters of a Gaussian distribution that are used to sample the next action to be applied in the POMDP. We refer to these transformed inputs as \emph{observations}, and we introduce the observation space $\mathcal{O}$. We optimise the parameter of one of these functions to maximise its expected return with PPO \cite{ppo}. Furthermore, we rely on an additional differentiable parameterised function, called the \emph{critic}, which we use in combination with PPO. We respectively denote the two functions as $\pi_\theta$ and $v_\phi$, where $\theta$ is the parameter of the function $\pi$ acting as the policy and $\phi$ is the parameter of the critic function $v$. The policy obtained after the last iteration of PPO is what we call a RL policy. After initialising the parameters $\theta$ and $\phi$, this algorithm iterates on the following steps (over a fixed number of iterations). It runs the policy over several independent simulations of the POMDP over a fixed time horizon $T$. The result of these simulations are sequences of transitions that are associated with reward signals. We refer to these sequences as \emph{episodes}. From these episodes, stored in a training set that we denote as $\mathcal{TS}$, it performs $N_{\text{upd}} \in \mathbb{N}_+$ updates of the policy $\pi_\theta$ (with respect to its parameters). For each update step $1 \leqslant i \leqslant N_{\text{upd}}$, a (small) subset of the transitions of size $BS \in \mathbb{N}_+$ is sampled, which is denoted as $TS_{i} \subseteq \mathcal{TS}$. Then, for each of these subsets, the update of the policy is computed as follows. Let 
\begin{equation}
\hat{A}_{\phi}(o_t) = \sum_{t'=t}^{T}{\left(\gamma'\lambda_{\text{GAE}}\right)^{t'-t} \left(r_t + \gamma v_{\phi'}(o_t) - v_{\phi}(o_t)\right)}
\end{equation}
\noindent be the so-called \emph{generalised advantage estimation} of the policy $\pi_\theta$ \cite{gae} where $v_{\phi'}$ and $v_{\phi}$ are respectively the prior and the current critic functions, $\gamma' \leqslant \gamma$ is the discount factor used in the PPO algorithm (that can be set lower than the one fixed for the POMDP \cite{franccois2015discount}) and $\lambda_{\text{GAE}} \in ]0; 1]$ is a hyperparameter that further increases the importance of the first reward signals. The descending gradient used to update the policy is computed from the following loss function:

\begin{align}
\label{eq:loss_func_ppo}
&\mathcal{L}^\pi (\theta) = - \frac{1}{BS} \sum_{(o_t, u_t, r_t) \in \mathcal{TS}} \\ \nonumber &\left[ \min( r_{\theta}(o_t, u_t) \hat{A}_{\phi}(o_t), \max(1 - \epsilon, \min(r_{\theta}(o_t, u_t), 1 + \epsilon)) \hat{A}_{\phi}(o_t)) \right],
\end{align}

\noindent where $\theta' = \theta$ before the first update, $r_{\theta}(o_t, u_t) = \frac{\pi_\theta(u_t \vert o_t)}{\pi_{\theta'}(u_t \vert o_t)}$ is the probability ratio between the initial policy and the updated one and $\epsilon$ is a hyperparameter that somehow limits the value of the loss function. The PPO algorithm updates the critic through another loss function which penalises the mean squared error of the critic function: 
\begin{equation}
\mathcal{L}^{v}(\phi) = \frac{\lambda_{v_{\phi}}}{BS} \sum_{(o_t, u_t, r_t) \in \mathcal{TS}}{\left(v_{\phi'}(o_t) - v_{\phi}(o_t)\right)^2},
\end{equation} where $\lambda_{v_{\phi}}$ is a hyperparameter that scales the importance of this loss function for updating the critic function. The gradient ascent updates are computed with ADAM \cite{kingma2014adam}, which dynamically modifies the learning rate, starting from an initial one to which we refer as $\eta$. These updates are usually clipped by a global norm defined by a hyperparameter that we denote as $\beta > 0$.

We implement parameterised policy and critic functions with deep neural networks, containing recurrent layers \cite{recurrentppo, yu2019review} to memorise the past exogenous variables in the form of hidden states. These hidden states are incorporated to the observation of the policies. We use the TBPTT algorithm \cite{williams1990efficient} to backpropagate the gradients of the recurrent layers up to a limited time horizon in the past transitions that we denote as $T_{\text{rnn}}$.

\newpage
\subsection{Baseline policies}
\label{math_baseline_policies}
The \emph{REC policy} and the \emph{SELF policy} compute the next action by respectively maximising the global self-consumption rate of the REC (as if the whole REC is composed of one member only) and the individual self-consumption rate for each member equipped with controllable assets. To that end, they respectively solve the optimisation problem defined below with $c_1 = 1$ and $c_2 = 1$, and setting the other coefficient to $0$:
\begin{mini!}|s|<b>
{\substack{u^{c, *}_{m, t} \in U(s^c_{m, t}),\\ \forall m \in \left\{1, \ldots, M\right\}}}{c_1 l^{REC}_{t} + c_2 \sum_{m \in \mathcal{M}_c}{ l_{m, t} }}
{}{}\label{eq:objreccons}
\addConstraint{l_{m, t}}{\geqslant l^-_{m, t} - l^+_{m, t}}{\forall m \in \left\{1, \ldots, M \right\}}
\addConstraint{l_{m, t}}{\geqslant l^+_{m, t} - l^-_{m, t}}{\forall m \in \left\{1, \ldots, M \right\}}
\addConstraint{l^{REC}_{t}}{\geqslant \sum_{m=1}^{M} l^-_{m, t} - l^+_{m, t}}{}
\addConstraint{l^{REC}_{t}}{\geqslant -\sum_{m=1}^{M} l^-_{m, t} - l^+_{m, t}}{}.
\end{mini!}
\noindent The terms enabled by the coefficient value $c1$ refer to the global self-consumption rate of the REC and the terms enabled by the coefficient value $c2$ refer to the individual self-consumption rates.  


\section{Illustrative examples of the optimal reallocation scheme}
\label{opt_scheme_example}
In this section, we provide three illustrative examples of the optimal reallocation scheme described in Section \ref{opt_realloc_scheme}. The first one is an REC composed of two members. The second one is an REC composed of three members, with peak cost coefficients set to $0$. The third one is another REC composed of three members with large peak costs (compared than the other cost coefficients). We consider, for all these RECs, a single billing period with two market periods. Note that, expected for the second REC, peak cost coefficients are always (significantly) larger than the other ones (prices of energy and distributions fees).

\subsection{REC composed of two members}
When there are only two members in the REC, that we denote as $m_1$ and $m_2$, the optimal solution can be computed directly as follows for each meter reading $r$. When one of the members has a positive net production and the other one has a positive net consumption (i.e., that $C^+_{m_1, r} > 0$ or $C^-_{m_2, r} > 0$ or vice versa), that member allocates as much as possible of its net production to the REC, and the REC reallocates the whole net electricity production to the other member. In other situations, no REC production is reallocated. In this REC scenario, either buying energy from the retailers instead of buying it from the REC if possible or selling energy instead of sharing it with the REC if possible is always suboptimal in both cases due to the assumptions on the inputs of this optimal reallocation scheme problem. Table \ref{tab:simple_rec_2_example_results} shows an example of an optimal reallocation scheme in this REC. 

\begin{table}
    \begin{subtable}{1.0\textwidth}
        \centering
\scalebox{0.85}{
\begin{tabular}{rr rr rrrr}
    \toprule
\multicolumn{2}{c}{\shortstack{Buying \\ \si{\text{\euro}/kWh}}} & \multicolumn{2}{c}{\shortstack{Selling \\ \si{\text{\euro}/kWh}}} &  \multicolumn{4}{c}{\shortstack{Network \\ \si{\text{\euro}/kWh}}}   \\
    \cmidrule(lr){0-1} \cmidrule(lr){3-4} \cmidrule(lr){5-8} 
        $M_1$  & $M_2$  & $M_1$  & $M_2$ & $P^-$ & $P^+$ & $\Lambda^-$ & $\Lambda^+$     \\ 
    \cmidrule(lr){0-1} \cmidrule(lr){3-4} \cmidrule(lr){5-8} 
        $\mathbf{0.20}$ & $\mathbf{0.22}$    & $\mathbf{0.04}$ & $\mathbf{0.05}$ & $\mathbf{1.00}$ & $\mathbf{1.00}$ & $\mathbf{0.02}$ & $\mathbf{0.03}$  \\
    \bottomrule
\end{tabular}
}
       
       \caption{}
       \label{tab:simple_rec_2_example_inputs}
    \end{subtable}
    \newline
\vspace*{0.05 cm}

\begin{subtable}{1.0\textwidth}
        \centering
\scalebox{0.85}{
\begin{tabular}{cc rr rr rr}
    \toprule
\multicolumn{2}{c}{} & \multicolumn{2}{c}{\shortstack{Net consumption \\ \si{kWh}}} & \multicolumn{2}{c}{\shortstack{Retail \\ \si{kWh}}} & \multicolumn{2}{c}{\shortstack{REC \\ \si{kWh}}} \\
    \cmidrule(lr){0-1} \cmidrule(lr){3-4} \cmidrule(lr){5-6} \cmidrule(lr){7-8} 
       \multicolumn{2}{c}{Market period} & $M_1$  & $M_2$ & $M_1$  & $M_2$ & $M_1$  & $M_2$      \\ 
        \cmidrule(lr){0-1} \cmidrule(lr){3-4} \cmidrule(lr){5-6} \cmidrule(lr){7-8} 
        \multicolumn{2}{c}{1} & $\mathbf{252.59}$ & $\mathbf{-596.18}$ & $\mathbf{0.00}$ & $\mathbf{-343.59}$ & $\mathbf{252.59}$ & $\mathbf{-252.59}$ \\
        \cmidrule(lr){0-1} \cmidrule(lr){3-4} \cmidrule(lr){5-6} \cmidrule(lr){7-8}

         \multicolumn{2}{c}{2} & $\mathbf{811.43}$ & $\mathbf{-244.02}$ & $\mathbf{567.41}$ & $\mathbf{0.00}$ & $\mathbf{244.02}$ & $\mathbf{-244.02}$ \\
        \cmidrule(lr){0-1} \cmidrule(lr){3-4} \cmidrule(lr){5-6} \cmidrule(lr){7-8}
        \multicolumn{2}{c}{} & \multicolumn{2}{c}{\shortstack{Offtake peak  \\ \si{kWh}}} & \multicolumn{2}{c}{\shortstack{Injection peak \\ \si{kWh}}} & \multicolumn{2}{c}{\shortstack{Global REC Bill \\ \si{\text{\euro}}}} \\
        \midrule 
       \multicolumn{2}{c}{Billing period} & $M_1$  & $M_2$  & $M_1$  & $M_2$  & NO-REC  & REC      \\ 
        \cmidrule(lr){0-1} \cmidrule(lr){3-4} \cmidrule(lr){5-6} \cmidrule(lr){7-8}
         \multicolumn{2}{c}{1} & $\mathbf{567.41}$  & $\mathbf{0.00}$   & $\mathbf{0.00}$ & $\mathbf{343.59}$  & \color{red}{$\mathbf{1578.40}$}  & \color{blue}{$\mathbf{1032.13}$}    \\ 
    \bottomrule
\end{tabular}
}
        
        \caption{}
        \label{tab:simple_rec_2_example_bills}
     \end{subtable}

\vspace*{0.05 cm}

     \caption{Optimal reallocation scheme problem over two market periods for an REC composed of two members. Retail prices and network costs are specified in (\subref{tab:simple_rec_2_example_inputs}). Energy exchanges, peaks and electricity bills are reported in (\subref{tab:simple_rec_2_example_bills}). The sum payable of the individual electricity bills (before accounting for energy exchanges through the REC) is reported in red. The global REC bill (after accounting for energy exchanges through the REC) is reported in blue.}
     \label{tab:simple_rec_2_example_results}
\end{table}

\subsection{REC without peaks costs}
When the offtake and injection peaks costs coefficients are set to zero, the optimal reallocation scheme can be solved with greedy algorithms for each market period \cite{dantzig_greedy}. More precisely, this greedy algorithm works as follows. Firstly, this algorithm identifies whether the net consumption (of the REC) is greater than the net production. If that is the case, it allocates the whole net production to the members in decreasing order of their energy buying prices. Otherwise, in a similar manner, it shares the net production of the members to the REC in the increasing order of their energy selling prices. Table \ref{tab:no_peak_costs_rec_3_example_results} shows the optimal reallocation scheme following that REC composition (i.e., an REC with three members where offtake and injection peak cost coefficients are set to $0$). Note that repartition schemes that are computed by the above-mentioned greedy algorithm are optimal only if all buying and selling prices are strictly positive.

\begin{table}
    \begin{subtable}{1.0\textwidth}
        \centering
\scalebox{0.85}{
\begin{tabular}{rrr rrr rrrr}
    \toprule
\multicolumn{3}{c}{\shortstack{Buying \\ \si{\text{\euro}/kWh}}} & \multicolumn{3}{c}{\shortstack{Selling \\ \si{\text{\euro}/kWh}}} &  \multicolumn{4}{c}{\shortstack{Network \\ \si{\text{\euro}/kWh}}}   \\
    \cmidrule(lr){0-2} \cmidrule(lr){4-6} \cmidrule(lr){7-10} 
        $M_1$  & $M_2$ & $M_3$  & $M_1$  & $M_2$ & $M_3$ & $P^-$ & $P^+$ & $\Lambda^-$ & $\Lambda^+$     \\ 
    \cmidrule(lr){0-2} \cmidrule(lr){4-6} \cmidrule(lr){7-10} 
        $\mathbf{0.20}$ & $\mathbf{0.22}$ & $\mathbf{0.24}$  & $\mathbf{0.04}$ & $\mathbf{0.05}$ & $\mathbf{0.06}$ & $\mathbf{0.00}$ & $\mathbf{0.00}$ & $\mathbf{0.02}$ & $\mathbf{0.03}$  \\
    \bottomrule
\end{tabular}
}     
       \caption{}
       \label{tab:no_peak_costs_rec_3_example_inputs}
    \end{subtable}
    \newline
\vspace*{0.05 cm}

\begin{subtable}{1.0\textwidth}
        \centering
\scalebox{0.65}{
\begin{tabular}{ccc rrr rrr rrr}
\toprule
\multicolumn{3}{c}{} & \multicolumn{3}{c}{\shortstack{Net consumption \\ \si{kWh}}} & \multicolumn{3}{c}{\shortstack{Retail \\ \si{kWh}}} & \multicolumn{3}{c}{\shortstack{REC \\ \si{kWh}}} \\
    \cmidrule(lr){0-2} \cmidrule(lr){4-6} \cmidrule(lr){7-9} \cmidrule(lr){10-12}
       \multicolumn{3}{c}{Market period} & $M_1$  & $M_2$ & $M_3$ & $M_1$  & $M_2$ & $M_3$ & $M_1$  & $M_2$ & $M_3$     \\ 
        \cmidrule(lr){0-2} \cmidrule(lr){4-6} \cmidrule(lr){7-9} \cmidrule(lr){10-12}
        \multicolumn{3}{c}{1} & $\mathbf{368.10}$ & $\mathbf{-608.36}$ & $\mathbf{-564.67}$ & $\mathbf{0.00}$ & $\mathbf{-240.26}$ & $\mathbf{0.00}$ & $\mathbf{368.10}$ & $\mathbf{-368.10}$ & $\mathbf{0.00}$ \\
        \cmidrule(lr){0-2} \cmidrule(lr){4-6} \cmidrule(lr){7-9} \cmidrule(lr){10-12}
        \multicolumn{3}{c}{2} &  $\mathbf{486.34}$ & $\mathbf{186.40}$ & $\mathbf{-162.35}$ & $\mathbf{486.34}$ & $\mathbf{24.05}$ & $\mathbf{-162.35}$ & $\mathbf{0.00}$ & $\mathbf{162.35}$ & $\mathbf{-162.35}$ \\
        \midrule
        \multicolumn{3}{c}{Billing period} &  $M_1$  & $M_2$ & $M_3$  & $M_1$  & $M_2$ & $M_3$  & NO-REC  & \multicolumn{2}{r}{REC}      \\ 
        \cmidrule(lr){0-2} \cmidrule(lr){4-6} \cmidrule(lr){7-9} \cmidrule(lr){10-12}
        \multicolumn{3}{c}{1} & /  & /   & /   & / & / & /  & \color{red}{$\mathbf{137.86}$}  & \multicolumn{2}{r}{\color{blue}{$\mathbf{83.19}$}}      \\ 
    \bottomrule
\end{tabular}
        }
        
        \caption{}
        \label{tab:no_peak_costs_rec_3_example_bills}
     \end{subtable}

\vspace*{0.05 cm}

     \caption{Optimal reallocation scheme over two market periods for an REC composed of three members which are not subject to peak costs. Retail prices and network costs are specified in (\subref{tab:no_peak_costs_rec_3_example_inputs}). Energy exchanges, peaks and electricity bills are reported in (\subref{tab:no_peak_costs_rec_3_example_bills}). The sum payable of the individual electricity bills (before accounting for energy exchanges through the REC) is reported in red. The global REC bill (after accounting for energy exchanges through the REC) is reported in blue.}
     \label{tab:no_peak_costs_rec_3_example_results}
\end{table}

\subsection{REC with peaks costs}

Despite our efforts, we did not identify any closed-form solution of the optimal reallocation scheme problem. However, as this optimisation problem is repeatedly solved with different inputs, we have observed during our tests that reusing the previous solutions significantly speeds-up the solving process. Table \ref{tab:general_rec_3_example_results_subopt_vs_opt} shows results of optimal reallocation schemes that either ignore or account for the peak costs.

\begin{table}
    
    \begin{subtable}{1.0\textwidth}
        \centering
    \scalebox{0.85}{
        \begin{tabular}{rrr rrr rrrr}
    \toprule
\multicolumn{3}{c}{\shortstack{Buying \\ \si{\text{\euro}/kWh}}} & \multicolumn{3}{c}{\shortstack{Selling \\ \si{\text{\euro}/kWh}}} &  \multicolumn{4}{c}{\shortstack{Network \\ \si{\text{\euro}/kWh}}}   \\
    \cmidrule(lr){0-2} \cmidrule(lr){4-6} \cmidrule(lr){7-10} 
        $M_1$  & $M_2$ & $M_3$  & $M_1$  & $M_2$ & $M_3$ & $P^-$ & $P^+$ & $\Lambda^-$ & $\Lambda^+$     \\ 
    \cmidrule(lr){0-2} \cmidrule(lr){4-6} \cmidrule(lr){7-10} 
        $\mathbf{0.20}$ & $\mathbf{0.22}$ & $\mathbf{0.24}$  & $\mathbf{0.04}$ & $\mathbf{0.05}$ & $\mathbf{0.06}$ & $\mathbf{1.00}$ & $\mathbf{1.00}$ & $\mathbf{0.02}$ & $\mathbf{0.03}$  \\
    \bottomrule
\end{tabular}
}
        \caption{}
        \label{tab:general_rec_3_example_bills_inputs_subopt_vs_opt}
     \end{subtable}

\begin{subtable}{1.0\textwidth}
\centering
\scalebox{0.65}{
\centering\begin{tabular}{ccc rrr rrr rrr}
\toprule

\multicolumn{3}{c}{} & \multicolumn{3}{c}{\shortstack{Net consumption \\ \si{kWh}}} & \multicolumn{3}{c}{\shortstack{Retail \\ \si{kWh}}} & \multicolumn{3}{c}{\shortstack{REC \\ \si{kWh}}} \\
    \cmidrule(lr){0-2} \cmidrule(lr){4-6} \cmidrule(lr){7-9} \cmidrule(lr){10-12}
        \multicolumn{3}{c}{Market period} & $M_1$  & $M_2$ & $M_3$ & $M_1$  & $M_2$ & $M_3$ & $M_1$  & $M_2$ & $M_3$     \\ 

        \cmidrule(lr){0-2} \cmidrule(lr){4-6} \cmidrule(lr){7-9} \cmidrule(lr){10-12}
        \multicolumn{3}{c}{1} & $\mathbf{-642.66}$ & $\mathbf{644.85}$ & $\mathbf{748.11}$ & $\mathbf{0.00}$ & $\mathbf{644.85}$ & $\mathbf{105.44}$ & $\mathbf{-642.66}$ & $\mathbf{0.00}$ & $\mathbf{642.66}$ \\

        \cmidrule(lr){0-2} \cmidrule(lr){4-6} \cmidrule(lr){7-9} \cmidrule(lr){10-12}
        \multicolumn{3}{c}{2} & $\mathbf{-666.00}$ & $\mathbf{142.05}$ & $\mathbf{-150.40}$ & $\mathbf{-523.94}$ & $\mathbf{0.00}$ & $\mathbf{-150.40}$ & $\mathbf{-142.05}$ & $\mathbf{142.05}$ & $\mathbf{0.00}$ \\

        \cmidrule(lr){0-2} \cmidrule(lr){4-6} \cmidrule(lr){7-9} \cmidrule(lr){10-12}
        \multicolumn{3}{c}{3} & $\mathbf{232.98}$ & $\mathbf{-111.48}$ & $\mathbf{813.45}$ & $\mathbf{232.98}$ & $\mathbf{0.00}$ & $\mathbf{701.97}$ & $\mathbf{0.00}$ & $\mathbf{-111.48}$ & $\mathbf{111.48}$ \\

        \cmidrule(lr){0-2} \cmidrule(lr){4-6} \cmidrule(lr){7-9} \cmidrule(lr){10-12}
        \multicolumn{3}{c}{4} & $\mathbf{-538.31}$ & $\mathbf{542.80}$ & $\mathbf{-579.49}$ & $\mathbf{0.00}$ & $\mathbf{0.00}$ & $\mathbf{-575.00}$ & $\mathbf{-538.31}$ & $\mathbf{542.80}$ & $\mathbf{-4.50}$ \\

        \midrule
        \multicolumn{3}{c}{} & \multicolumn{3}{c}{\shortstack{Offtake peak  \\ \si{kWh}}} & \multicolumn{3}{c}{\shortstack{Injection peak \\ \si{kWh}}} & \multicolumn{3}{c}{\shortstack{Global REC Bill \\ \si{\text{\euro}}}} \\
        \cmidrule(lr){0-2} \cmidrule(lr){4-6} \cmidrule(lr){7-9} \cmidrule(lr){10-12} 
        \multicolumn{3}{c}{Billing period} & $M_1$  & $M_2$ & $M_3$  & $M_1$  & $M_2$ & $M_3$  & NO-REC  & \multicolumn{2}{r}{REC}      \\ 
        \cmidrule(lr){0-2} \cmidrule(lr){4-6} \cmidrule(lr){7-9} \cmidrule(lr){10-12}
        \multicolumn{3}{c}{1} & $\mathbf{232.98}$  & $\mathbf{644.85}$   & $\mathbf{701.97}$   & $\mathbf{523.94}$ & $\mathbf{0.00}$ & $\mathbf{575.00}$  & \color{red}{$\mathbf{3638.90}$}  & \multicolumn{2}{r}{\color{purple}{$\mathbf{3068.45}$}}      \\ 
    \bottomrule
\end{tabular}
}
        
        \caption{}
        \label{tab:general_rec_3_example_bills_subopt}
     \end{subtable}
\begin{subtable}{1.0\textwidth}
\centering
\scalebox{0.65}{
\centering\begin{tabular}{ccc rrr rrr rrr}
\toprule

\multicolumn{3}{c}{} & \multicolumn{3}{c}{\shortstack{Net consumption \\ \si{kWh}}} & \multicolumn{3}{c}{\shortstack{Retail \\ \si{kWh}}} & \multicolumn{3}{c}{\shortstack{REC \\ \si{kWh}}} \\
    \cmidrule(lr){0-2} \cmidrule(lr){4-6} \cmidrule(lr){7-9} \cmidrule(lr){10-12}
        \multicolumn{3}{c}{Market period} & $M_1$  & $M_2$ & $M_3$ & $M_1$  & $M_2$ & $M_3$ & $M_1$  & $M_2$ & $M_3$     \\ 

        \cmidrule(lr){0-2} \cmidrule(lr){4-6} \cmidrule(lr){7-9} \cmidrule(lr){10-12}
        \multicolumn{3}{c}{1} & $\mathbf{-642.66}$ & $\mathbf{644.85}$ & $\mathbf{748.11}$ & $\mathbf{0.00}$ & $\mathbf{2.18}$ & $\mathbf{748.11}$ & $\mathbf{-642.66}$ & $\mathbf{642.66}$ & $\mathbf{0.00}$ \\

        \cmidrule(lr){0-2} \cmidrule(lr){4-6} \cmidrule(lr){7-9} \cmidrule(lr){10-12}
        \multicolumn{3}{c}{2} & $\mathbf{-666.00}$ & $\mathbf{142.005}$ & $\mathbf{-150.4}$ & $\mathbf{-523.94}$ & $\mathbf{0.00}$ & $\mathbf{-150.4}$ & $\mathbf{-142.05}$ & $\mathbf{142.05}$ & $\mathbf{0.00}$ \\

        \cmidrule(lr){0-2} \cmidrule(lr){4-6} \cmidrule(lr){7-9} \cmidrule(lr){10-12}
        \multicolumn{3}{c}{3} & $\mathbf{232.98}$ & $\mathbf{-111.48}$ & $\mathbf{813.45}$ & $\mathbf{186.85}$ & $\mathbf{0.00}$ & $\mathbf{748.11}$ & $\mathbf{46.13}$ & $\mathbf{-111.48}$ & $\mathbf{65.34}$ \\

        \cmidrule(lr){0-2} \cmidrule(lr){4-6} \cmidrule(lr){7-9} \cmidrule(lr){10-12}
        \multicolumn{3}{c}{4} & $\mathbf{-538.31}$ & $\mathbf{542.8}$ & $\mathbf{-579.49}$ & $\mathbf{-424.59}$ & $\mathbf{0.00}$ & $\mathbf{-150.4}$ & $\mathbf{-113.71}$ & $\mathbf{542.8}$ & $\mathbf{-429.09}$ \\

        \midrule
        \multicolumn{3}{c}{} & \multicolumn{3}{c}{\shortstack{Offtake peak  \\ \si{kWh}}} & \multicolumn{3}{c}{\shortstack{Injection peak \\ \si{kWh}}} & \multicolumn{3}{c}{\shortstack{Global REC Bill \\ \si{\text{\euro}}}} \\
        \cmidrule(lr){0-2} \cmidrule(lr){4-6} \cmidrule(lr){7-9} \cmidrule(lr){10-12} 
        \multicolumn{3}{c}{Billing period} & $M_1$  & $M_2$ & $M_3$  & $M_1$  & $M_2$ & $M_3$  & NO-REC  & \multicolumn{2}{r}{REC}      \\ 
        \cmidrule(lr){0-2} \cmidrule(lr){4-6} \cmidrule(lr){7-9} \cmidrule(lr){10-12}
        \multicolumn{3}{c}{1} & $\mathbf{186.85}$  & $\mathbf{2.18}$   & $\mathbf{748.11}$   & $\mathbf{523.94}$ & $\mathbf{0.00}$ & $\mathbf{150.4}$  & \color{red}{$\mathbf{3638.9}$}  & \multicolumn{2}{r}{\color{blue}{$\mathbf{2024.38}$}}      \\ 
    \bottomrule
\end{tabular}}
        
        \caption{}
        \label{tab:general_rec_3_example_bills_opt}
     \end{subtable}
    \newline
\vspace*{-0.75 cm}
\newline
     \caption{Optimal reallocation scheme over four market periods for an REC composed of four members. Retail prices and network costs are specified in (\subref{tab:general_rec_3_example_bills_inputs_subopt_vs_opt}). Energy exchanges, peaks and electricity bills (that results from ignoring the peak costs) are reported in (\subref{tab:general_rec_3_example_bills_subopt}). The sum payable of the individual electricity bills (before accounting for energy exchanges through the REC) is reported in red. The global REC bills (after accounting for energy exchanges through the REC) computed by ignoring and accounting for the peak costs are reported in purple and blue, respectively.}
     \label{tab:general_rec_3_example_results_subopt_vs_opt}
\end{table}

\section{Experimental protocol details}
\label{rlpoliciesdetails}
In this section, we provide the details of the experimental protocol (including numerical values) under which the policies, described in detail in Appendix \ref{math_policies}, have been simulated in REC-2 and REC-7. More precisely, we outline the configuration on the exogenous variables sampling for both REC instances, as well as the structure of each REC instance, and the configuration of the MPC and RL policies with respect to these REC instances.

\subsection{Sampling exogenous variables}
\label{sample_exo_algo}
Weather- and daytime-related time series, notably solar-based energy production and the companies' energy consumption, are typically time correlated. In an attempt to replicate these time series dynamics, we propose the following approach to simulate $P_0^{\mathcal{E}}$ and $P^{\mathcal{E}}$ from historical data. Let $e$ be an exogenous time series. Let $\mathbf{\omega}$ be a white noise time series (centred on zero) sampled with a standard deviation that we denote as $\sigma$. To take into account time correlation, we derive a new noise $x$, to which we refer as \emph{red noise}\footnote{\url{https://atmos.washington.edu/~breth/classes/AM582/lect/lect8-notes.pdf}}, as follows:
\begin{align}
    x_0 &= \omega_0,\\
    x_{t+1} &=  rx_t + \sqrt{(1-r^2)}w_{t+1}, \; \forall t \geqslant 1,
\end{align} \noindent where $r \in \; ]0,1]$ controls the time correlation of the red noise. We then define the two distributions $P_0^{\mathcal{E}}$ and $P^{\mathcal{E}}$ such that the value of a sample $\tilde{e}_t$ of the two distributions is given at time step $t$ by
\begin{align}
    \tilde{e}_t = x_t + e_t.
\end{align}

\subsection{Future exogenous variable values for MPC policies}
\label{mpc_predicting_algo}
Let $\tilde{e}$ be the historical sequence of exogenous variables from which new sequences are sampled through the procedure described in Section \ref{sample_exo_algo}. Let $e$ be the exogenous time series containing the future values from the time step $t$ (as sampled through the procedure described in Section \ref{sample_exo_algo}). We thus define the time series $\hat{e}_{t:t+K}$ to be the prediction of the exogenous variables from $t$ to $t+K$, computed as follows:

\begin{equation}
    \label{eq:pseudo_forecast}
    \hat{e}_t' = \alpha^{t' - t}e_{t'} + (1 - \alpha^{t' - t})\tilde{e}_{t'}, \; \forall t' \in \left\{t, \ldots, t+K \right\},
\end{equation} \noindent where $\alpha$ controls the convergence speed of the exogenous time series to the mode of the two distributions $P^{\mathcal{E}}_0$ and $P^{\mathcal{E}}$. Note that, in practice, these predictions are computed through algorithms built with supervised learning techniques \cite{cons_forecast, prod_forecast}, provided that a reasonably large training dataset is available. This is not the case for the context of our research work, where the scarcity of historical data for RECs challenges the building process of such forecasting algorithms.

\subsection{REC-2}
\label{rec_2_details}
\begin{figure}
\centering
\includegraphics[width=14.5cm]{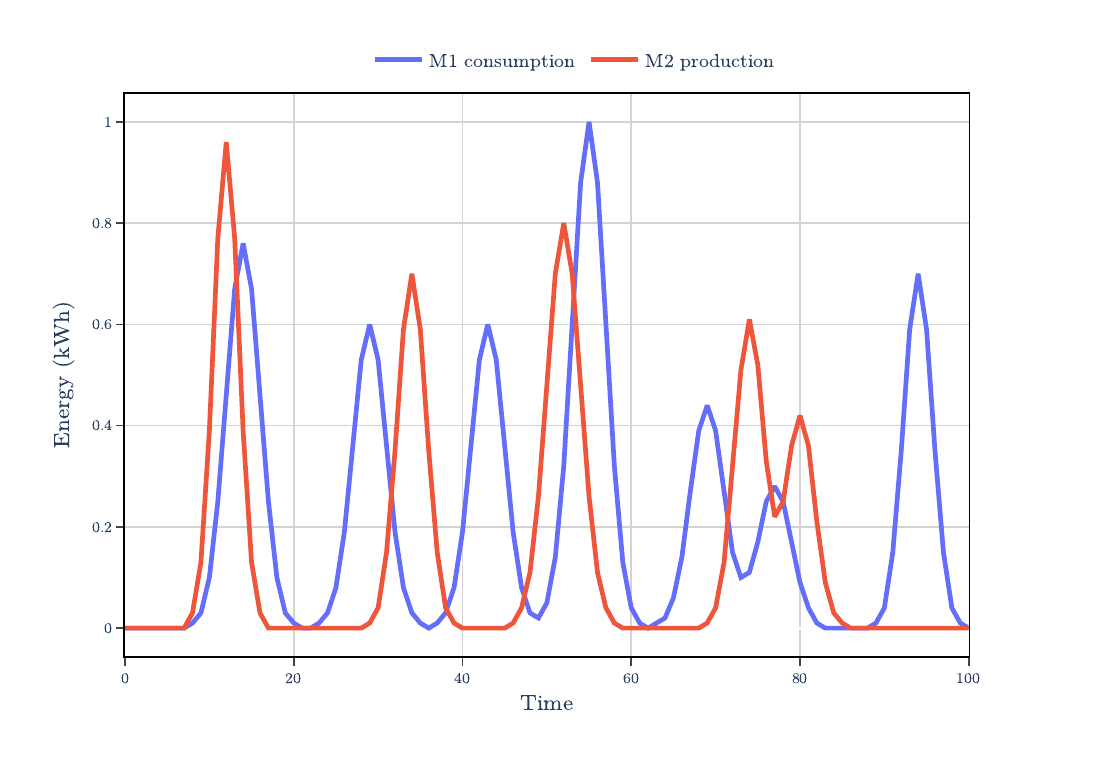}
\vspace{-0.75cm}
\caption{Consumption/production profiles of members in REC-2. Member M1 does not produce electricity and member M2 does not consume electricity (through their respective non-controllable assets).}
\label{fig:rec2profiles}
\end{figure}

REC-2 is composed of a consumer and a producer, which we denote as \emph{M1} and \emph{M2}, respectively; M1 is equipped with non controllable consuming devices only, and M2 is equipped with non controllable consuming and producing devices, such that it (M2) produces more energy than it consumes. The duration between two time steps $\delta$ is set to one hour. A billing period occurs every five market periods. A market period lasts four time steps. In other words, optimal reallocation schemes are computed for the last $20$ hours at each end of any billing period. Figure \ref{fig:rec2profiles} shows the consumption and production profiles of the members. Table \ref{tab:retail_prices_rec2} shows the buying and selling prices as defined by retailer contracts in REC-2. Offtake and injection peak cost coefficients are both set to $1$ \si{\text{\euro}/kWh}. Network fees are set to $\Lambda^- = 0.03$ \si{\text{\euro}/kWh} and $\Lambda^+ = 0.01$ \si{\text{\euro}/kWh}. The member M2 is equipped with a battery following linear charging power dynamics. More formally, let $s^c_{M2}$ be the state of charge of the battery, let $u_{M2}$ be the charging power applied to the battery (negative is discharging), let $U^{-}$ and $U^{+}$ be the respective discharging and charging powers bounds, and let $S^c_{M2}$ be its maximum capacity. The dynamics of the state of charge is defined as follows:

\begin{align}
\label{eq:battery_details}
s^c_{M2, 0} &= \frac{S^c_{M2}}{2},\\ \nonumber
s^c_{M2, t+1} &= s^c_{M2, t} + \delta \left[\nu^+ u^+_{M2, t} - \frac{u^-_{M2, t}}{\nu^-} \right],
\end{align}

\noindent where $u^+_{M2, t}$ and $u^-_{M2, t}$ are respectively charging and discharging battery power, and $\nu^+$ and $\nu^-$ are respectively charging and discharging efficiencies. Table \ref{tab:battery_specs_rec2} shows the producer's battery specifications. The time horizon of the simulations is fixed to $101$ time steps, with a discount factor of $0.9995$. 

\begin{table}
\centering
\begin{tabular}{crr}
    \toprule
    \shortstack{REC \\ Member} & \shortstack{Buying \\ \si{\text{\euro}/kWh}} & \shortstack{Selling \\ \si{\text{\euro}/kWh}}\\
    \midrule
    M1 & $0.10$ & $0.01$ \\
    M2 & $0.12$ & $0.01$ \\
    \bottomrule
\end{tabular}
\vspace{0.33cm}
\caption{Buying and selling prices of members in REC-2.}
    \label{tab:retail_prices_rec2}
\end{table}

\begin{table}
\centering
\begin{tabular}{crr}
    \toprule
    Specification & \multicolumn{2}{c}{Value}\\
    \midrule
    Maximum capacity ($S^c_{M2}$) & $1.00$ & \si{kWh} \\
    Maximum charging power ($U^+$) & $0.05$ & \si{kW} \\
    Maximum discharging power ($U^-$) & $0.10$ & \si{kW} \\
    Charging efficiency ($\nu^+$) & $1.00$ & \\
    Discharging efficiency ($\nu^-$) & $1.00$ & \\
    \bottomrule
\end{tabular}\\
\vspace{0.33cm}
\caption{Specifications of the battery owned by member M2 in REC-2.}
    \label{tab:battery_specs_rec2}
\end{table}

\subsubsection{Exogenous variables sampling}

Exogenous variables are sampled from the consumption and production profiles of the REC members, as shown in Figure \ref{fig:rec2profiles}. To generate the red noise, we set the correlation parameter to $r=0.5$ and the standard deviation of the white noise to $\sigma=0.3$.

\subsubsection{MPC and baseline policies}

To evaluate the MPC and the baseline policies, we sample $64$ exogenous time series through the procedure described in Appendix \ref{sample_exo_algo}, and we run the simulations with each of these time series independently. More precisely, they are evaluated by averaging the sample expected returns resulting from these simulations through $16$ distinct random seeds. We run the MPC policies with the values of $\alpha$ in $\left\{ 1.0, 0.85, 0.5 \right\}$ and the values of $K \in \left\{1, 2, \ldots, 101\right\}$.

\subsubsection{RL policies}
The RL policies transform input data to its corresponding observation (as specified in Appendix \ref{math_rl_policies}) as follows. The states of the controllable assets and the counters (as continuous values) are kept. For each member, the values of the consumption(production) meter readings and the net consumption(production) -- before the controllable assets usage -- are summed up for the current time step; the other values (of exogenous variables and meter readings) of the previous time steps are discarded. Afterwards, the consumption and production meter reading are replaced, for each member, by a net consumption meter reading (a negative value a net production meter reading). Since standardisation of the observation (and the rewards) is recommended for training policies in a reinforcement learning setting \cite{whatmattersonpolicy}, we proceed to that end as follows. Let $O$ be an observation vector of size $N > 1$. Each value $o_i$ of this observation vector, with $i$ between $1$ and $N$, is shifted by a fixed mean value and divided by a fixed standard deviation value. These statistics are computed by a simulation of the OPT policy (described in Section \ref{baselinepol}) through the historical exogenous variables. Note that, for the RL dense policies, we add to the observation the last reward computed during the current billing period. These rewards are standardised in the same fashion as the observations. 

\begin{figure}
\centering
\hspace*{1.5cm}\includegraphics[scale=0.8]{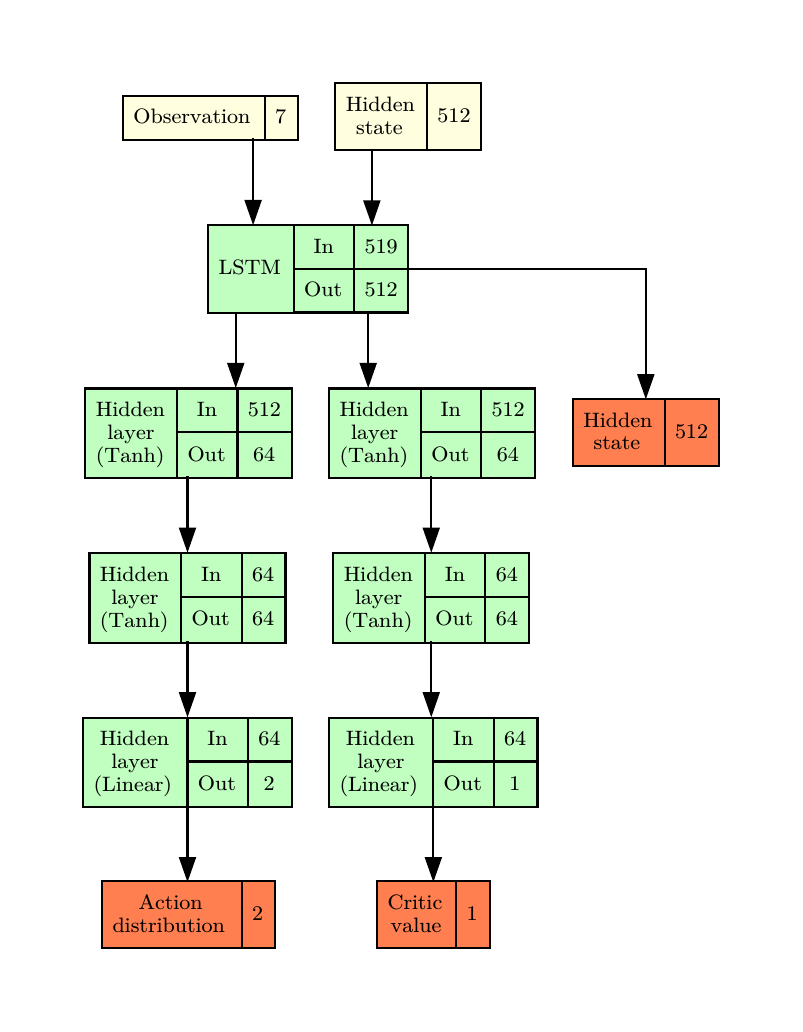}
\vspace{-0.75cm}
\caption{Forward view of the underlying deep recurrent learning model of RL policies for REC-2 for a single observation. Yellow nodes are the inputs (including the RNN hidden states). Green nodes are hidden layers with activation functions. Orange nodes are outputs (including the next RNN hidden states).}
\label{fig:rec2neuralnetwork}
\end{figure}

Thereafter, the observation is fed to the underlying deep neural network, for which the architecture is shown in Figure \ref{fig:rec2neuralnetwork}. The input hidden state is either the last output of the recurrent layers, or an initial vector of zero values (before the first step of an episode). The result of this forward computation is a pair of values corresponding to the parameters of a Gaussian distribution, namely the mean and the log standard deviation. Before sampling a value from this distribution, the standard deviation is transformed by the exponential function and shifted with a value $\epsilon = 1e^{-6}$ (to avoid exploding gradients). The sampled value is then clipped between $-1$ and $1$ and projected into the bounds of the action space (in this case, between the maximum discharging and the maximum charging powers of the battery). If this action is not admissible (e.g., the discharging battery power is greater than the content of the battery), it is projected to the closest feasible value (e.g., discharging power value corresponding to the content of the battery).

The weights of the underlying neural network of the prior policy (before the first update of the PPO algorithm) are initialised accordingly to the programming library RLLib \cite{rllib}. Additionally, as recommended by \cite{whatmattersonpolicy}, we divide the weights of the last layer outputting the action distribution by a hyperparameter that we denote as $W$. Through the simulation process of policy functions, the training set records the transitions along with the respective parameters of the (Gaussian) distribution that sampled the actions and the critic values. It also records the standardised values of the rewards, which are used instead of the original reward signals to compute the loss function of the policy.  The policy is then updated as explained in Appendix \ref{math_policies}. Table \ref{tab:rl_policy_hyperopt_rec2} shows the hyperparameters (including the grid search space) used to train the RL policies.

\begin{figure}
\centering
\includegraphics[width=15cm]{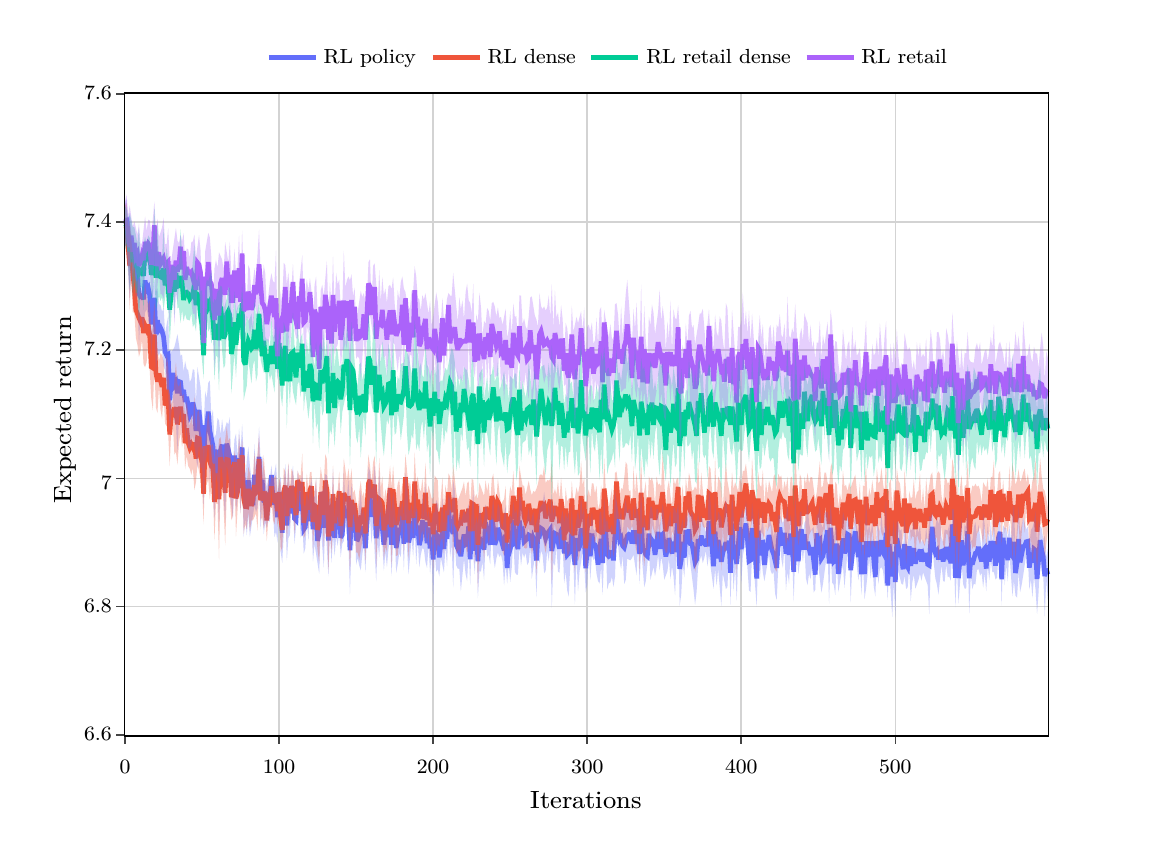}
\vspace{-0.5cm}
\caption{Mean expected return with standard error over training iterations of the RL policies in REC-2.}
\label{fig:rec2rlresults}
\end{figure}

To evaluate the RL policies, we ran the PPO algorithm for $600$ iterations through $16$ distinct random seeds. For each iteration, the current policy was run through $64$ simulations in the environment (by sampling $64$ exogenous time series). At the end of this procedure, $64$ additional simulations were sampled, and the expected return estimate was computed by averaging the expected return samples across the random seeds. Figure \ref{fig:rec2rlresults} shows the evaluation of the RL policies at each iteration of the PPO algorithm. 

\begin{table}
\centering
\scalebox{0.8}{
\begin{tabular}{crrrrr}
    \toprule
    Hyperparameter & Search space & RL policy & RL dense & RL retail & RL retail dense\\
    \midrule
    $\lambda_{\text{GAE}}$ & $\left\{0.90, 0.95, 0.99\right\}$ & $0.90$ & $0.90$ & $0.90$ & $0.90$\\
    $\eta$ & $\left\{5\mathrm{e}{-4}, 5\mathrm{e}{-5}, 5\mathrm{e}{-6} \right\}$ & $5\mathrm{e}{-5}$ & $5\mathrm{e}{-5}$ & $5\mathrm{e}{-5}$ & $5\mathrm{e}{-5}$\\
    $\gamma$ & $\left\{0.9995, 0.95, 0.99\right\}$ & $0.99$ & $0.99$ & $0.99$ & $0.99$ \\
    $\beta$ & $\left\{1, 2, 4\right\}$ & $2$ & $2$ & $1$ & $1$ \\
    $N_{\text{upd}}$ & $\left\{5, 10\right\}$ & $10$ & $10$ & $10$ & $10$\\
    $B$ & $\left\{32, 64, 96\right\}$ & $64$ & $64$ & $64$ & $64$ \\
    $\lambda_{v_{\phi}}$ & $\left\{1, 0.1, 0.01, 0.001  \right\}$ & $1$ & $0.01$ & $1$ & $1$ \\
    $T_{\text{rnn}}$ & $\left\{25, 50, 100\right\}$ & $50$ & $50$ & $50$ & $50$ \\
    $W$ & $\left\{1, 10, 100\right\}$ & $1$ & $1$ & $1$ & $1$ \\
    \bottomrule
\end{tabular}}\\
\vspace{0.33cm}
\caption{Values of hyperparameters related to PPO for RL policies in REC-2, with the search space in the second column; see Appendix \ref{math_policies} for more details about the hyperparameters. Unspecified hyperparameters have been left to default values defined by the programming library RLlib \cite{rllib}.}
    \label{tab:rl_policy_hyperopt_rec2}
\end{table}

\subsection{REC-7}
\label{rec_7_details}

\begin{figure}
\centering
\includegraphics[scale=0.8]{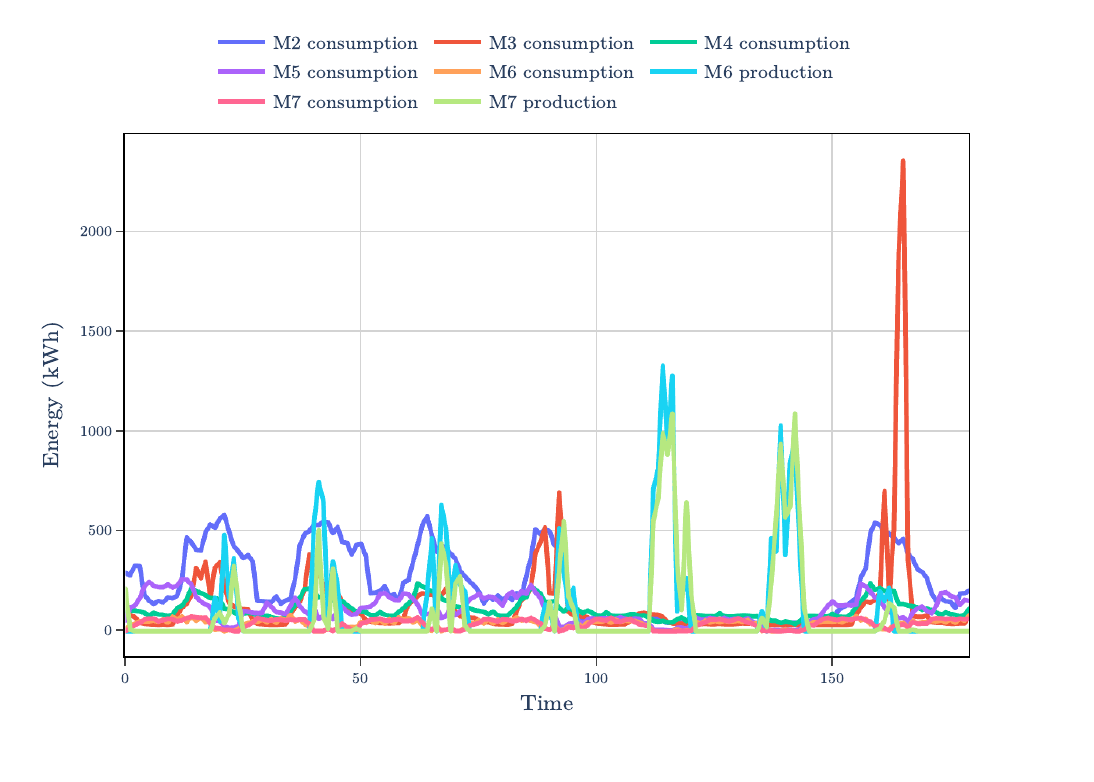}
\vspace{-0.75cm}
\caption{Consumption and production profiles for REC-7 (resampled for ease of readability). Members M2, M3, M4 and M5 do not produce electricity. Members M6 and M7 both consume and produce electricity. Electricity flows from member M1 are only generated from their battery.}
\label{fig:rec7profiles}
\end{figure}

From now on, as the experimental settings share many similarities with those in REC-2, we only describe the parts that differ from the latter. REC-7 is composed of seven members, which are denoted from \emph{M1} to \emph{M7}. The duration between two time steps $\delta$ is set to three minutes. A billing period occurs every $45$ market periods. Figure \ref{fig:rec7profiles} shows the consumption and production profiles of the members,  derived from historical data of an existing REC in Wallonia, Belgium. Table \ref{tab:retail_prices_rec7} shows the buying and selling prices by the members as defined by their retailer contracts. Offtake and injection peak cost coefficients are both set to $1.210$ \si{\text{\euro}/kWh}. Network fees are set to $\Lambda^- = 0.143$ \si{\text{\euro}/kWh} and $\Lambda^+ = 0.126$ \si{\text{\euro}/kWh}. Similarly to REC-2, only member M1 is equipped with a battery following the same charging dynamics as described by Equation \eqref{eq:battery_details}. Table \ref{tab:battery_specs_rec7} shows the producer's battery specifications (by reusing the notation introduced in Section \ref{rec_2_details}). The time horizon of the simulations is fixed to $720$ time steps, with a discount factor of $0.99993$. 

\begin{table}
\centering
\begin{tabular}{crr}
    \toprule
    \shortstack{REC\\ Member} & \shortstack{Buying\\\si{\text{\euro}/kWh}} & \shortstack{Selling\\\si{\text{\euro}/kWh}}\\
    \midrule
    M1 & $0.214907$ & $0.075388$\\
    M2 & $0.208757$ & $0.075152$\\
    M3 & $0.202735$ & $0.076381$\\
    M4 & $0.20846$ & $0.077213$ \\
    M5 & $0.20846$ & $0.078153$ \\
    M6 & $0.206301$ & $0.080649$ \\
    M7 & $0.210234$ & $0.081928$ \\
    \bottomrule
\end{tabular}\\
\vspace{0.33cm}
\caption{Retail buying and selling prices for REC-7.}
    \label{tab:retail_prices_rec7}
\end{table}

\begin{table}
\centering
\begin{tabular}{crr}
        \toprule
    Specification & \multicolumn{2}{c}{Value}\\
    \midrule
    Maximum capacity ($S^c_{M2}$) & $5256$ &  \si{kWh}  \\
    Maximum charge power ($U^+$) & $525$ & \si{kW}  \\
    Maximum discharge power ($U^-$) & $1051$ & \si{kW}  \\
    Charging efficiency ($\nu^+$) & $0.88$ & \\
    Discharging efficiency ($\nu^-$) & $0.71$ &  \\
    \bottomrule
\end{tabular}\\
\vspace{0.33cm}
\caption{Specifications of the battery of member M1 in REC-7.}
    \label{tab:battery_specs_rec7}
\end{table}

\subsubsection{MPC and baseline policies}
To evaluate the MPC and the baseline policies, we sample $64$ exogenous time series through the procedure described in Appendix \ref{sample_exo_algo}, and we run the simulations with each of these time series independently. More precisely, they are evaluated by averaging the sample of expected returns resulting from these simulations through $16$ distinct random seeds. We run the MPC policies with the values of $\alpha$ in $\left\{ 1.0, 0.95, 0.5 \right\}$ and the values of $K$ in $\left\{ 1, \ldots, 100, 132, 164, 196, 266, 330, 394, 458, 522, 586, 650, 721\right\}$.

\subsubsection{RL policies}

\begin{figure}
\centering
\includegraphics[scale=0.8]{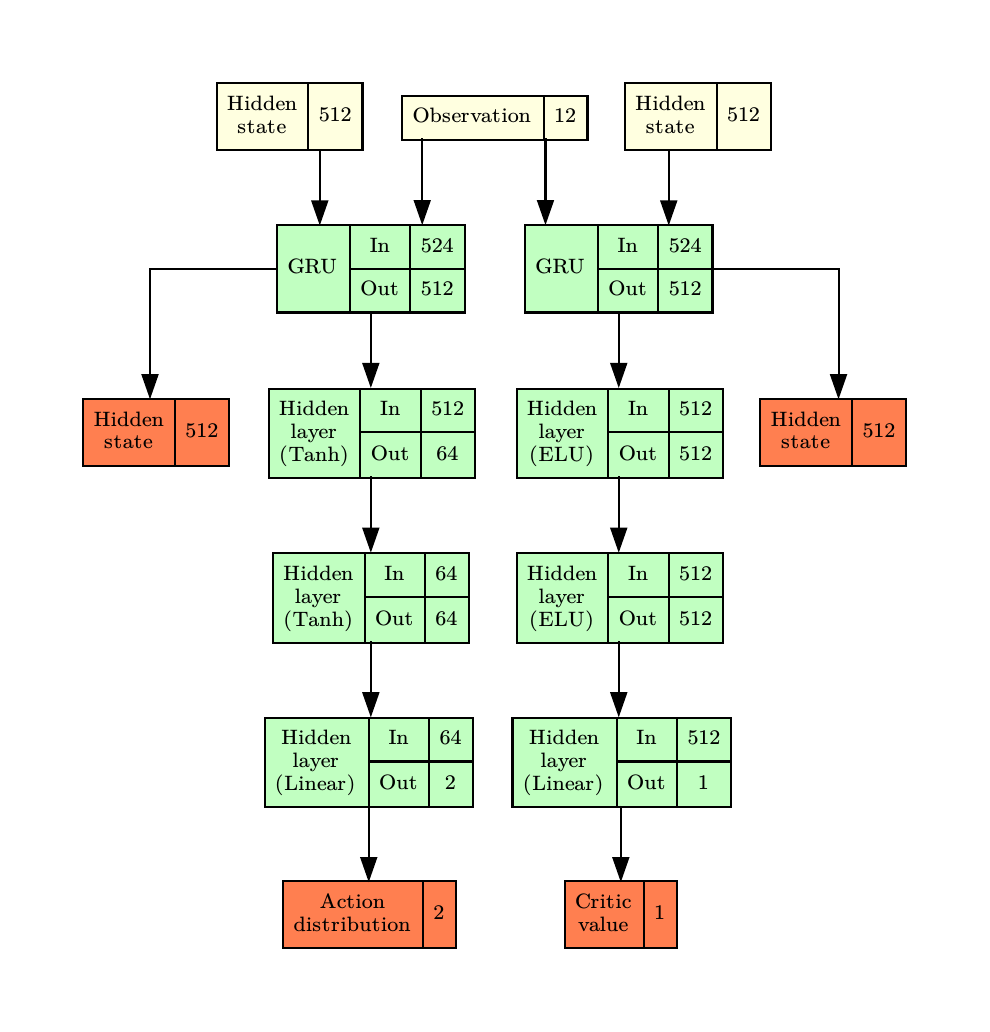}
\vspace{-0.75cm}
\caption{Forward view of underlying deep recurrent learning model of RL policies for REC-7 for a single observation. Yellow nodes are the inputs (including the RNN hidden states). Green nodes are hidden layers with activation functions. Orange nodes are output (including the next RNN hidden states).}
\label{fig:rec7neuralnetwork}
\end{figure}

\begin{figure}
\centering
\includegraphics[width=14.5cm]{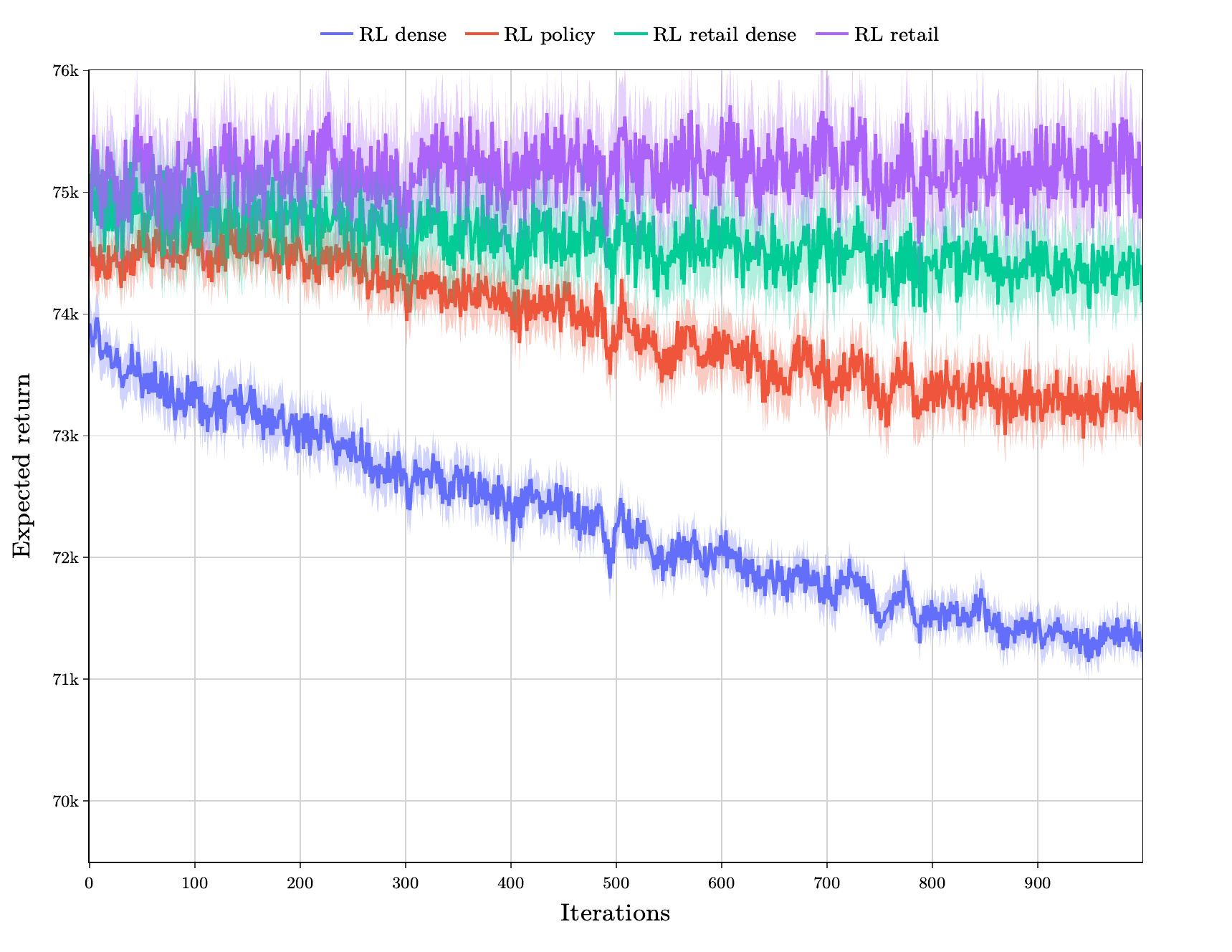}
\vspace{-0.75cm}
\caption{Mean expected return with standard error over training iterations of the RL policies in REC-7.}
\label{fig:rec7rlresults}
\end{figure}

Figure \ref{fig:rec7neuralnetwork} shows the architectures of two independent deep neural networks. The first one computes the parameters of Gaussian distributions (to sample the next actions) and the second one is the critic value. Table \ref{tab:rl_policy_hyperopt_rec7} shows the hyperparameters (including the grid search space) used to train the RL policies. To evaluate the RL policies, we ran the PPO algorithm for $1000$ iterations through $16$ distinct random seeds. For each iteration, the current policy was run through $128$ simulations in the environment. At the end of this procedure, $128$ additional simulations were sampled to estimate the expected return. Figure \ref{fig:rec7rlresults} shows the evaluation of the RL policies at each iteration of the PPO algorithm. 


\begin{table}
\centering
\scalebox{0.8}{
\begin{tabular}{ccrrrr}
    \toprule
    Hyperparameter & Search space & RL policy & RL dense & RL retail & RL retail dense\\
    \midrule
    $\lambda_{\text{GAE}}$ & $\left\{0.9, 0.95, 0.99\right\}$ & $0.9$ & $0.9$ & $0.9$ & $0.9$\\
    $\eta$ & $\left\{5\mathrm{e}{-4}, 5\mathrm{e}{-5}, 5\mathrm{e}{-6} \right\}$ & $5\mathrm{e}{-5}$ & $5\mathrm{e}{-5}$ & $5\mathrm{e}{-5}$ & $5\mathrm{e}{-5}$\\
    $\gamma$ & $\left\{0.99993, 0.95, 0.99\right\}$ & $0.99$ & $0.99$ & $0.99$ & $0.99$ \\
    $\beta$ & $\left\{1, 2, 4\right\}$ & $2$ & $2$ & $1$ & $1$ \\
    $N_{\text{upd}}$ & $\left\{5, 10\right\}$ & $5$ & $5$ & $5$ & $5$\\
    $B$ & $\left\{360, 720\right\}$ & $360$ & $360$ & $360$ & $360$ \\
    $\lambda_{v_{\phi}}$ & $\left\{1, 0.1, 0.01, 0.001  \right\}$ & $0.001$ & $0.001$ & $0.001$ & $0.001$ \\
    $T_{\text{rnn}}$ & $\left\{180, 360\right\}$ & $360$ & $360$ & $360$ & $360$ \\
    $W$ & $\left\{1, 10, 100\right\}$ & $100$ & $100$ & $100$ & $100$ \\
    \bottomrule
\end{tabular}}\\
\vspace{0.33cm}
\caption{Values of hyperparameters related to PPO for RL policies in REC-7, with the search space in the second column; see Appendix \ref{math_policies} for more details about the hyperparameters. Unspecified hyperparameters have been left to default values defined by the programming library RLlib \cite{rllib}.}
    \label{tab:rl_policy_hyperopt_rec7}
\end{table}

\end{document}